\documentclass[preprint2]{aastex631}

\usepackage{longtable}

\def\dsr{$R_\odot$}

\def\dteff{$T_{\mathrm{eff}}$}
\def\dalpha{$\alpha_{\rm{MLT}}$}

\shorttitle{Rotating Solar Models}
\shortauthors{Wuming Yang}

\begin{document}


\title{Rotating Solar Models in Agreement with Helioseismic Results and Updated Neutrino Fluxes}
\author[0000-0002-3956-8061]{Wuming Yang}
\affiliation{Department of Astronomy, Beijing Normal University, Beijing 100875, China}
\email{yangwuming@bnu.edu.cn}

%\author{August Muench}
% \affiliation{American Astronomical Society \\
% 1667 K Street NW, Suite 800 \\
% Washington, DC 20006, USA}

\begin{abstract}
Standard solar models (SSMs) constructed in accordance with old solar abundances are in reasonable agreement
with seismically inferred results, but SSMs with new low-metal abundances disagree with the seismically inferred
results. The constraints of neutrino fluxes on solar models exist in parallel with those of helioseismic
results. The solar neutrino fluxes were updated by Borexino Collaboration. We constructed rotating solar models
with new low-metal abundances where the effects of enhanced diffusion and convection overshoot were included.
A rotating model using OPAL opacities and the Caffau abundance scale has better sound-speed and
density profiles than the SSM with the old solar abundances and reproduces the observed $p$-mode frequency
ratios $r_{02}$ and $r_{13}$. The depth and helium abundance of the convection zone of the model agree with
the seismically inferred ones at the level of $1\sigma$. The updated neutrino fluxes are also reproduced by the model
at the level of $1\sigma$. The effects of rotation and enhanced diffusion not only improve the model's
sound-speed and density profiles but bring the neutrino fluxes predicted by the model into agreement with the detected
ones. Moreover, the calculations show that OP may underestimate opacities for the regions of the Sun with
$T\gtrsim5\times10^{6}$ K by around $1.5\%$, while OPAL may underestimate opacities for the regions of the Sun
with $2\times10^{6}$ K $\lesssim T \lesssim 5\times10^{6}$ K by about $1-2\%$.
\end{abstract}

\section{Introduction}
The heavy-element abundance $Z_{s}$ of the Sun, derived by \citet[hereafter GS98]{grev98} from photospheric
spectroscopy, is $0.017$, whose uncertainty is of the order of $10$ percent. The ratio of the heavy-element
abundance to hydrogen abundance is $0.0231$. Since \citet{lod03} and \citet{aspl05} reassessed the value of
the $Z_{s}$, it has been revised several times \citep{lod09, aspl09, caf10, caf11, lod20, aspl21, amar21}.
The well-known values of the new $Z_{s}$ are $0.0133$ \citep{lod03}, $0.0122$ \citep{aspl05}, $0.0141$ \citep{lod09},
or $0.0134$ \citep[hereafter AGSS09]{aspl09}. These revised values are obviously lower than the old one.

The helium abundance, $Y_{s}$, in the solar convection zone (CZ) and thus photosphere cannot be inferred directly
from spectroscopy, but can be determined by helioseismology. The values of seismically inferred $Y_{s}$ and $Z_{s}$
are $0.2485\pm0.0035$ \citep{basu04, sere10} and $0.0172\pm0.002$ \citep{anti06}, respectively. However, the values
given by \citet{voro13} or \citet{voro14} are in the range of $Y_{s}=0.240–0.255$ and $Z_{s}=0.008–0.013$ or
$Y_{s}=0.245–0.260$ and $Z_{s}=0.006–0.011$. Moreover, the value of $Z_{s}$ inferred by \citet{buld17}
is in the range of $0.008–0.014$. The radius of the base of the CZ (BCZ) also can be determined by helioseismology.
The inferred radius of the BCZ is $0.713\pm0.003$ \dsr{} \citep{chr91} or $0.713\pm0.001$ \dsr{} \citep{basu97}.

The standard solar models (SSMs) constructed in accordance with the high metal abundances (old
solar abundances, e.g. GS98) are considered to be in good agreement with the seismically inferred
sound-speed and density profiles, depth and helium abundance of the CZ, but the SSMs constructed in accordance
with the low metal abundances (revised solar abundances) do not completely agree with the seismically
inferred results \citep{bah04b, basu04, yang07, basu09, sere09, sere11, zhang12} and the neutrino flux constraints
\citep{bah04a, tur10, tur11, tur11a, yang16}, which is known as solar modeling problem or solar abundance
problem \citep{basu15, chris21, salm21, amar21}.

In order to reconcile the low-Z models with helioseismology, many physical effects have been
studied. For example, increased opacity at the base of the CZ was studied by \citet{bah04b},
\citet{sere09}, and \citet{buld19}; enhanced neon abundance was suggested by \citet{bah05};
mass accretion of metal-rich/poor material or helium-poor material was investigated by \citet{cast07},
\citet{guzi10}, \citet{sere11}, and \citet{zhang19}; overshooting below the CZ was used to recover
the CZ depth \citep{mont06, cast07, yang19, zhang19}.

In order to match the seismically inferred sound-speed and density profiles \citep{basu00, basu09},
the gravitational settling that reduces the surface helium abundance by about $11\%$ ($\sim 0.03$
by mass fraction) below its initial value is required in SSMs. Macroscopic turbulent mixing can
reduce the amount of surface helium settling by around $40\%$ \citep{pro91}. The effects
of the turbulent mixing on the diffusion and settling of helium and heavy elements were not
considered in the diffusion coefficients of \citet{tho94}. \citet{aspl04} suggested that enhanced
diffusion and settling of helium and heavy elements might be able to reconcile the low-Z models
with helioseismology. The increased diffusion can significantly improve sound-speed and density
profiles, but leaves the CZ helium abundance too low \citep{basu04, mont04, guzi05, yang19}.

Rotational mixing can transport helium outward. It thus can counteract the effect of enhanced
diffusion on the surface helium abundance \citep{yang19}, i.e., it can improve the prediction
of the surface helium abundance. The effects of rotation on the low-Z models were studied by \cite{yang07},
\citet{tur10, tur11}, and \citet{yang16, yang19}. However, the rotating models with AGSS09 mixtures
\citep{yang19} disagree with the detected neutrino fluxes of \citet{bore18, bore20}.

\citet{bah04a} and \citet{sere09} found that an $11\%-20\%$ increase in OPAL opacities at the BCZ
can reconcile the low-Z models with helioseismology. \citet{badn05} showed that in the region OP opacity is
slightly larger than OPAL opacity but no more than $2.5$ percent. \citet{buld19} concluded that the solar
modeling problem likely occurs from multiple small contributors. Opacity could be one of the contributors.

The production of solar neutrinos is sensitive to the central properties of the Sun. The $^{8}$B neutrino flux
is strongly dependent on the central temperature of the Sun \citep{bah88, tur11a}. The determinations of solar
neutrino fluxes complement helioseismology in diagnosing the core of the Sun. The constraints of neutrino fluxes
on solar models exist in parallel with those of helioseismology and are studied by many authors \citep{bah04a, tur10,
tur11, tur11a, yang16, yang19, zhang19}. \citet{bore18, bore20} updated solar neutrino fluxes and determined
the total fluxes of $^{13}$N, $^{15}$O, and $^{17}$F neutrinos. Their $^{8}$B neutrino flux is higher than
that determined by \citet{berg16}. The low-Z model of \citet{zhang19} with helium-poor accretion and solar-wind
mass loss agrees with helioseismic results but disagrees with the neutrino fluxes detected by \citet{bore18, bore20}.
Moreover, \citet{salm21} showed that the updated solar neutrino fluxes prefer high-metallicity solar models.
The solar modeling problem has persisted for almost 20 years.

\citet{caf10, caf11} independently analysed carbon, nitrogen, and oxygen abundances in the solar photosphere.
They found that the heavy-element abundance of the solar surface is $Z_{s}=0.0154/0.0153$, supplemented with data
from \citet{lod09}. The value of $Z_{s}/X_{s}$ advocated by \citet{caf11} is $0.0209$. These values are larger than
those of \citet{lod09} and AGSS09. \citet{hope20} studied the possible origin of the solar abundance problem.
Their result favors the solar $Z_{s}$ reported by \citet{caf11} rather than the measurement by AGSS09.

\citet{lod20} reanalysed the solar photospheric abundances and recommended $Z_{s}=0.0149$ and $Z_{s}/X_{s}=0.0201$.
\citet[hereafter AAG21]{aspl21} also reassessed the solar abundances with updated atomic data and a 3D
radiative-hydrodynamical model of the solar photosphere. They advocated $Z_{s}=0.0139\pm0.0006$, $Y_{s}=0.2423\pm0.0054$,
and $Z_{s}/X_{s}=0.0187\pm0.0009$. Moreover, using the 3D radiative-hydrodynamical model of solar photosphere,
\citet{amar21} analysed $408$ molecular lines of C, N, and O and obtained solar C, N, and O abundances,
which are slightly larger than their previous results in AAG21. With these molecular abundances, the value of the
heavy-element abundance of the solar surface increases from $Z_{s}=0.0139$ to $Z_{s}=0.0142$ \citep{amar21}.
This increase indicates that ``it may be worthwhile to continue improving the atomic and molecular data as well as
the model atmospheres and line formation methods'' \citep{amar21}.

In this work, we mainly focus on whether the solar models constructed in accordance with AAG21 mixtures
and \citet{caf11} mixtures agree with the seismically inferred results and updated neutrino
fluxes. The paper is organized as follows. Input physics in models are introduced in Section 2,
calculation results are presented in Section 3, and discussion and summary are given in Section 4.

\section{Input Physics in Models}

All solar models were calculated by using the Yale Rotating Stellar Evolution Code \citep{pins89, yang07,
dema08} in its rotation and non-rotation configurations. The frequencies of $p$-modes of models were computed
by using the \citet{gue94} pulsation code. The OPAL equation-of-state (EOS2005) tables \citep{rog02},
OPAL \citep{igl96} and OP opacity tables \citep{seat87, op95, badn05, dela16} were used, supplemented
by the \citet{fer05} opacity tables at low temperature. The opacity tables were reconstructed with
the GS98, \citet{caf11}, and AAG21 mixtures (see Appendix \ref{appda}). The nuclear reaction
rates were calculated with the subroutine of \citet{bah92} and \citet{bah95, bah01} (see Appendix
\ref{appdb}). Convection was determined by the Schwarzschild criterion and treated according to the
standard mixing-length theory \citep{boh58, kipp12}. The overshoot region below the BCZ was assumed
to be both fully mixed and adiabatically stratified \citep{chr91}. The depth of the overshoot region is
determined by $\delta_{\rm ov}H_{p}$ \citep{dema08}, where $\delta_{\rm ov}$ is a free parameter and $H_{p}$ is
the local pressure scale height. A convection overshoot of $\delta_{\rm ov}\approx0.1$ is required to recover
the seismically inferred depth of the CZ in our rotating models. The diffusion and settling of both helium and
heavy elements were computed using the formulas of \citet{tho94}. In the atmosphere, \citet{kris66} $T-\tau$ relation
was adopted.

We treated the transport of angular momentum and material mixing as a diffusion process \citep{enda78},
i.e.
    \begin{equation}
       \frac{\partial \Omega}{\partial t}=f_{\Omega}
       \frac{1}{\rho r^{4}}\frac{\partial}{\partial r}(\rho r^{4}D
       \frac{\partial \Omega}{\partial r}) \,
      \label{diffu1}
    \end{equation}
for the transport of angular momentum and
    \begin{equation}
    \begin{array}{lll}
        \frac{\partial X_{i}}{\partial t}&=&f_{c}f_{\Omega}\frac{1}{\rho r^{2}}
       \frac{\partial}{\partial r}(\rho r^{2}D\frac{\partial X_{i}}
        {\partial r})\\
        & &+(\frac{\partial X_{i}}{\partial t})_{\rm nuc}-\frac{1}
       {\rho r^{2}}\frac{\partial}{\partial r}(f_{0}\rho r^{2}X_{i}V_{i}) \,
    \end{array}
      \label{diffu2}
    \end{equation}
for the change in the mass fraction $X_{i}$ of chemical species $i$, where $D$ is the diffusion coefficient
caused by rotational instabilities including the Eddington circulation, the Goldreich-Schubert-Fricke
instability \citep{pins89}, and the secular shear instability of \cite{zahn93}. The default values of $f_{\Omega}$
and $f_{c}$ are $1$ and $0.03$ \citep{yang19}, respectively. We applied a straight multiplier $f_{0}$ to
the diffusion velocity $V_{i}$ to enhance the rates of diffusion and settling, as \citet{basu04},
\citet{mont04}, \citet{guzi05}, and \citet{yang19} have done, despite the fact that there is no obvious
physical justification for such a multiplier. The value of $f_{0}$ is $1$ for standard cases but larger
than $1$ for an enhanced diffusion model. The angular-momentum loss from the CZ due to magnetic braking
was calculated with Kawaler’s relation \citep{kaw88, cha95}. More details of calculation for rotation were
described in \citet{pins89} and \citet{yang19}.

All models were evolved from a homogeneous zero-age main-sequence model to the present solar age $4.57$ Gyr,
luminosity $3.844\times10^{33}$ erg $\mathrm{s}^{-1}$, radius $6.9598\times10^{10}$ cm, and mass
$1.9891\times10^{33}$ g \citep{bah95}. The initial metallicity $Z_{0}$, hydrogen abundance $X_{0}$,
and mixing-length parameter \dalpha{} are free parameters. They were adjusted to match the constraints of
luminosity and radius within around $10^{-5}$ and observed $Z_{s}/X_{s}$. The initial helium abundance
is determined by $Y_{0}=1-X_{0}-Z_{0}$. The value of \dalpha{} changes with input physics, i.e. a new solar
calibration of \dalpha{} is performed each time the input physics change. The initial rotation rate, $\Omega_{i}$
of rotating models was adjusted to reproduce the solar equatorial velocity of about $2.0$ km s$^{-1}$.
The values of these parameters are shown in Table \ref{tab1}.

\section{Calculation Results}
\subsection{Solar Models with High Metal Abundances}
\subsubsection{The Models Constructed with OPAL Opacity Tables}
Using the OPAL opacity tables constructed with GS98 mixtures, we computed SSM GS98M and rotating
model GS98Mr. The central temperature and density, surface helium and heavy-element abundances,
radius of the BCZ, and other parameters of the models are listed in Table \ref{tab1}. The fluxes
of $pp$, $pep$, $hep$, $^{7}$Be, $^{8}$B, $^{13}$N, $^{15}$O, and $^{17}$F neutrinos calculated
from the models are given in Table \ref{tab2}. Table \ref{tab3} lists some of the physical
configurations of each model.

\begin{figure*}
\includegraphics[angle=-90, scale=0.8]{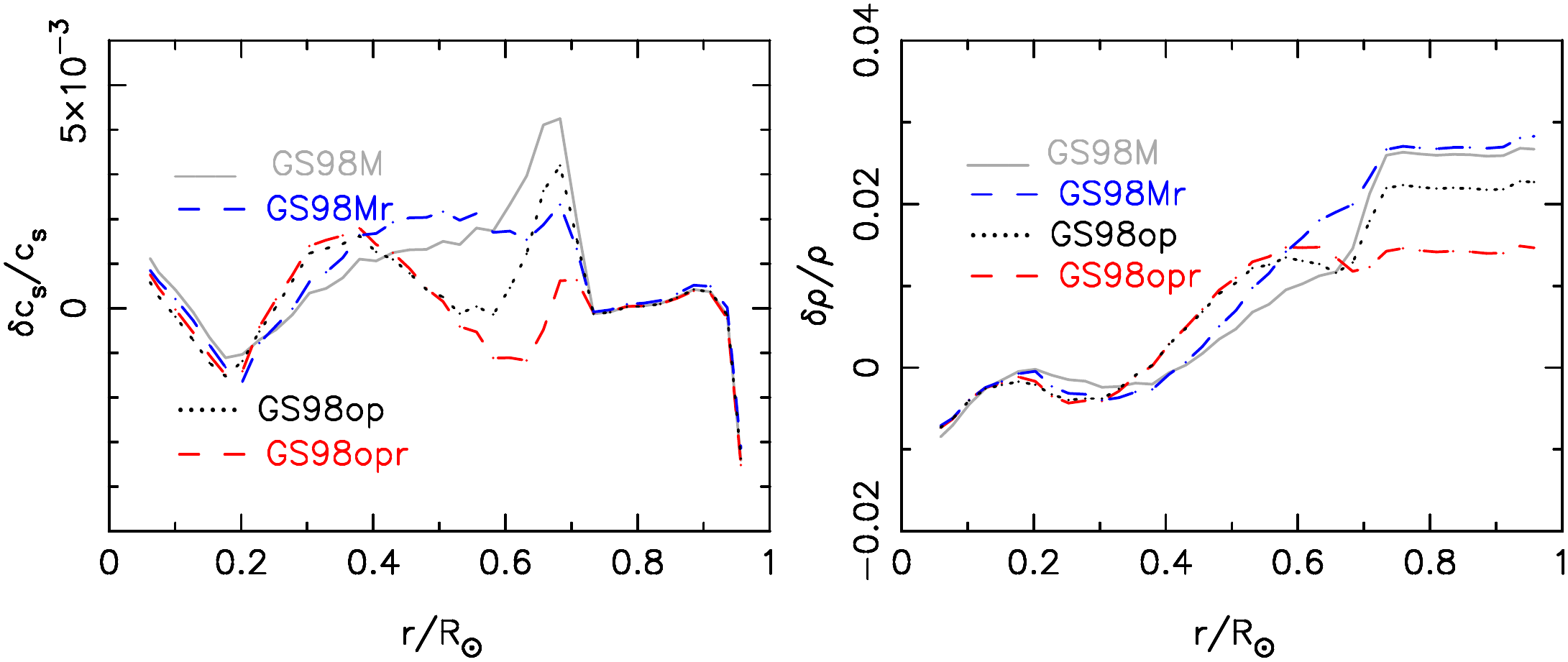}
\caption{Relative sound-speed and density differences, in the sense (Sun-Model)/Model, between the Sun
and models. The solar sound speed and density are given in \citet{basu09}.
\label{fig1}}
\end{figure*}

\begin{figure*}
\includegraphics[angle=-90, scale=0.8]{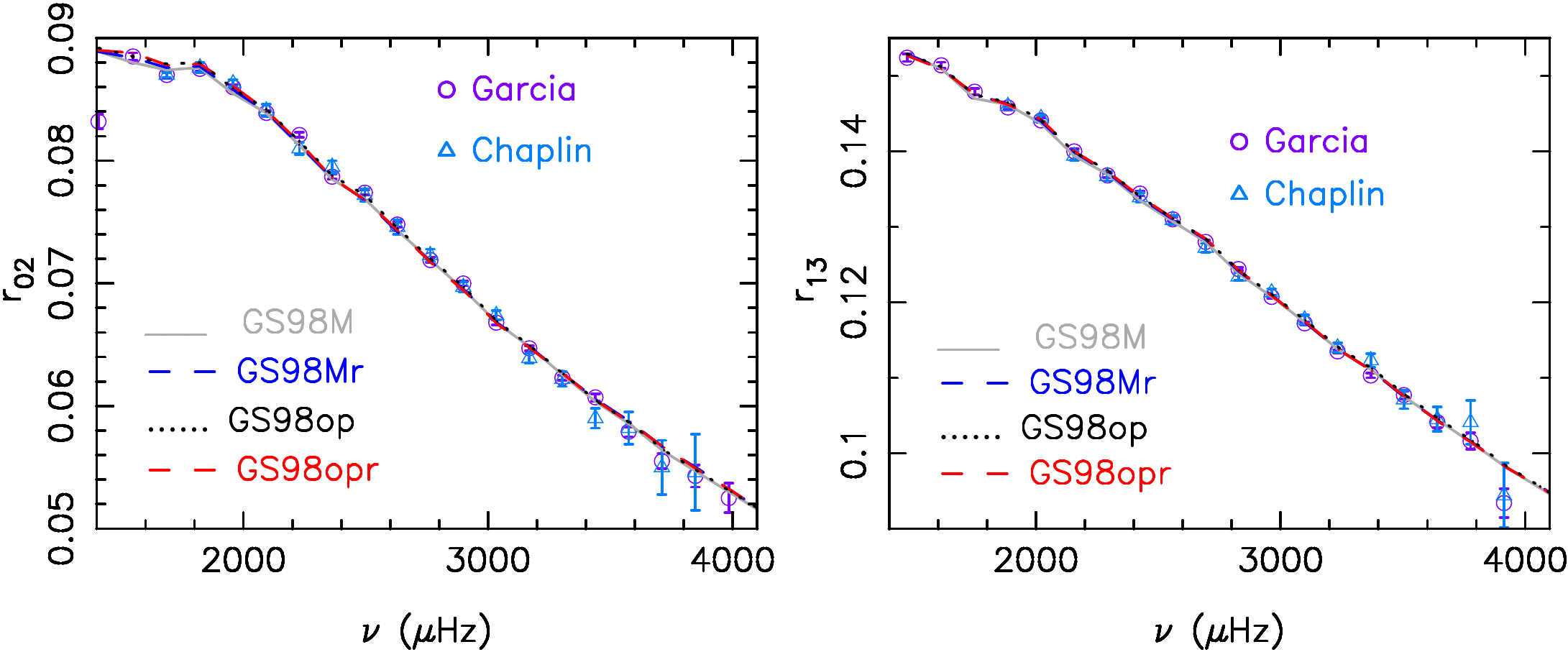}
\caption{Distributions of observed and predicted ratios $r_{02}$ and $r_{13}$ as a function
of frequency. The circles and triangles show the ratios calculated from the frequencies observed
by GOLF \& VIRGO \citep{gar11} and BiSON \citep{cha99b}, respectively.
\label{fig2}}
\end{figure*}

The surface heavy-element abundance of $0.0174$ of GS98M is in agreement with that determined by GS98.
The surface helium abundance and $r_{cz}$ of GS98M are also consistent with the seismically
inferred ones (see Table \ref{tab1}). We compared the sound speed and density of models
with those inferred by \citet{basu09} using the data from the Birmingham Solar-Oscillations
Network \citep{cha96} and the Michelson Doppler Imager \citep{sch98}. The values of relative
sound-speed difference, $\delta c_{s}/c_{s}$, and density difference, $\delta \rho/\rho$, between
the Sun and GS98M are less than $0.0043$ and $0.028$, respectively (see Figure \ref{fig1}). Moreover,
the ratios of small to large frequency separations, $r_{02}$ and $r_{13}$ \citep{rox03}, of the model agree
with those calculated from observed frequencies of \citet{cha99b} or \citet{gar11} (see Figure \ref{fig2}).
Table \ref{tab4} gives the values of $\chi^{2}_{c_{s}+\rho}$ and $\chi^{2}_{d_{02+13}}$
of the models.

We compared the neutrino fluxes computed from models with those determined by different
authors \citep{bel11, bel12, ahm04, berg16, bore18, bore20} and ones predicted by the models
BP04 \citep{bah04a} and SSeM \citep{tur11a} in Table \ref{tab2}. The neutrino fluxes of GS98M
are comparable with those of BP04. Some differences between the nuclear cross-section factors
$S_{0}$ \citep{bah89} used in BP04 and those used in GS98M are listed in Table \ref{tabnuc} of
Appendix \ref{appdb}. The total fluxes of $^{13}$N, $^{15}$O, and $^{17}$F
neutrinos calculated from GS98M are $\Phi(\rm CNO)=10.2\times10^{8}$ cm$^{-2}$ s$^{-1}$,
slightly larger than $7^{+3}_{-2}\times10^{8}$ cm$^{-2}$ s$^{-1}$ detected by
\citet{bore20}. The $pp$, $pep$, $hep$, $^{7}$Be, and $^{8}$B neutrino fluxes of GS98M
are in agreement with those determined by \citet{bore18}. However, the fluxes of $^{7}$Be
and $^{8}$B neutrinos are larger than those determined by \citet{berg16} (see Table \ref{tab2}).

Figure \ref{fig1} shows that the effects of rotation can significantly improve the sound-speed
and density profiles. The value of the relative sound-speed difference, $\delta c_{s}/c_{s}$,
below the CZ can be decreased by about $50\%$. Moreover, the centrifugal effect leads to a decrease
in the central temperature. The fluxes of $^{7}$Be, $^{8}$B, $^{13}$N, $^{15}$O, and $^{17}$F neutrinos
are sensitive to the central temperature. Thus the fluxes calculated from rotating models are generally
lower than those computed from non-rotating models (see Table \ref{tab2}). The total fluxes
of $^{13}$N, $^{15}$O, and $^{17}$F neutrinos of GS98Mr are $\Phi(\rm CNO)=9.5 \times10^{8}$
cm$^{-2}$ s$^{-1}$, which are in agreement with the detected value of $7^{+3}_{-2}\times10^{8}$
cm$^{-2}$ s$^{-1}$. The $^{8}$B neutrino flux of $5.29 \times10^{6}$ cm$^{-2}$ s$^{-1}$ for GS98Mr
is consistent with $5.16^{+0.13}_{-0.09} \times10^{6}$ cm$^{-2}$ s$^{-1}$ determined by \citet{berg16}
and $5.68^{+0.39}_{-0.41} \times10^{6}$ cm$^{-2}$ s$^{-1}$ detected by \citet{bore18}.
However, the surface helium abundance of 0.2534 of GS98Mr is higher than the inferred value of
$0.2485\pm0.0035$. Thus the high-Z models constructed with OPAL opacities do not completely agree
with helioseismic results and updated neutrino fluxes.

\subsubsection{The Models Constructed with OP Opacity Tables}
\begin{figure*}
\includegraphics[angle=0, scale=0.5 ]{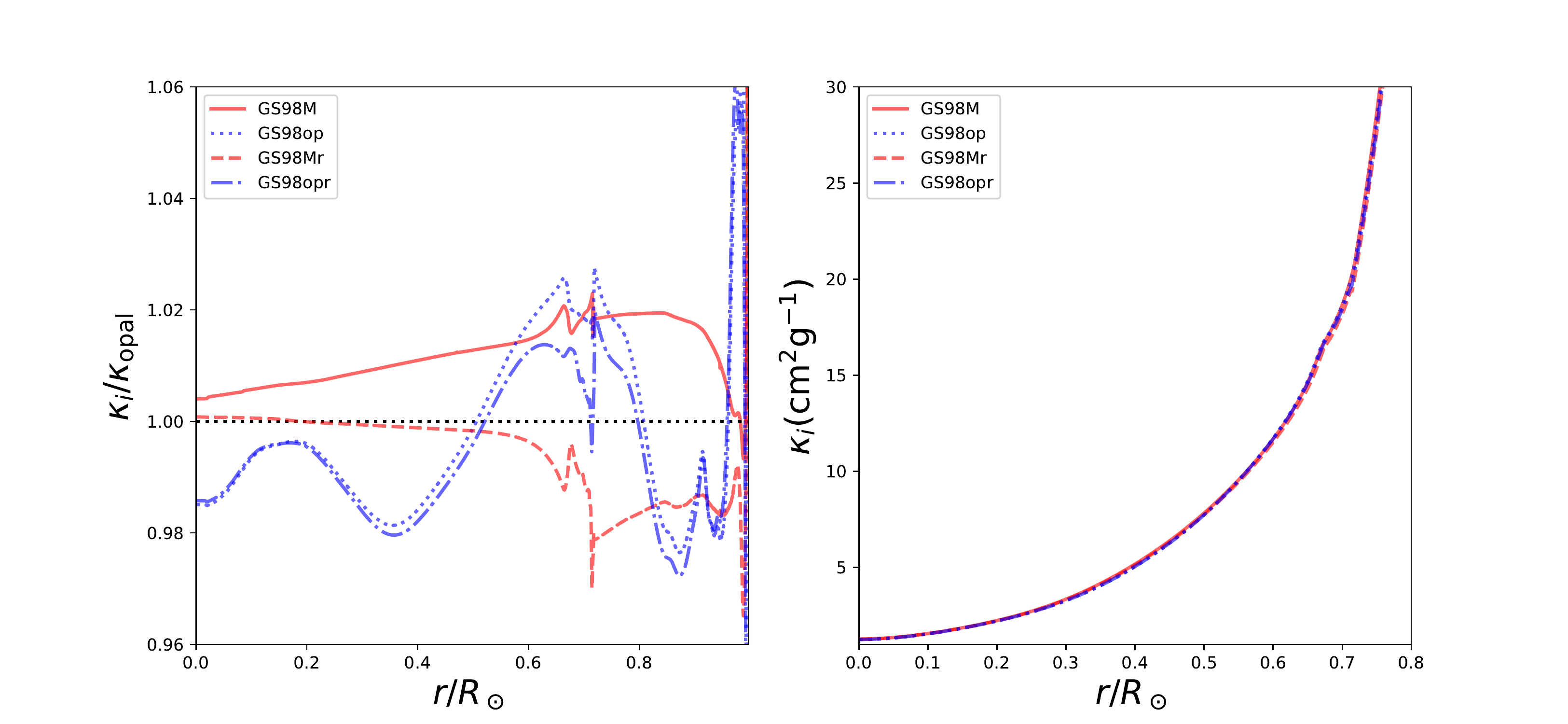}
\caption{Left panel: comparison of Rosseland mean opacity of different models relative to that of
a reference model constructed with OPAL opacity tables and $Z_{s}=0.0170$.
Right panel: distributions of Rosseland mean opacity of the models as a function of radius.
\label{fig3}}
\end{figure*}

In order to study the effects of opacities, we constructed SSM GS98op and rotating model GS98opr
by using OP opacity tables. Tables \ref{tab1} and \ref{tab2} list the fundamental parameters and
neutrino fluxes of the models, respectively. For almost the same metallicity, OP opacities are
larger by about $0-2\%$ than OPAL opacities in the region of the Sun with $0.5$ \dsr{}$\lesssim r \lesssim 0.8$ \dsr{}
but smaller by about $1-2\%$ in the region with $r \lesssim 0.4$ \dsr{} (see Figure \ref{fig3}).
The sound-speed and density profiles and frequency separation ratios of the models are
shown in Figures \ref{fig1} and \ref{fig2}, respectively, which show that OP has almost no
improvement in the reproduction of the frequency separation ratios in comparison
to OPAL (see Figure \ref{fig2}), but obviously improves density profile and the sound-speed
profile below the CZ where OP opacities are mainly larger than OPAL opacities. OP slightly
worsens the sound-speed profile in the inner layers of radiative region where
OP opacities are lower than OPAL opacities (see Figures \ref{fig1} and \ref{fig3}).

In order to obtain the same surface heavy-element abundance and the solar luminosity and radius at
the age of $4.57$ Gyr, the decrease in opacities requires decreasing the initial helium abundance and
changing \dalpha{}. As a consequence, the surface helium abundances of the models constructed with
OP opacity tables are lower than those of the models constructed with OPAL opacity tables. The surface helium
abundance of $0.2414$ for GS98op is smaller than the inferred value of $0.2485\pm0.0035$. Therefore,
GS98op is not a good SSM.

\begin{figure*}
\includegraphics[angle=0, scale=0.5 ]{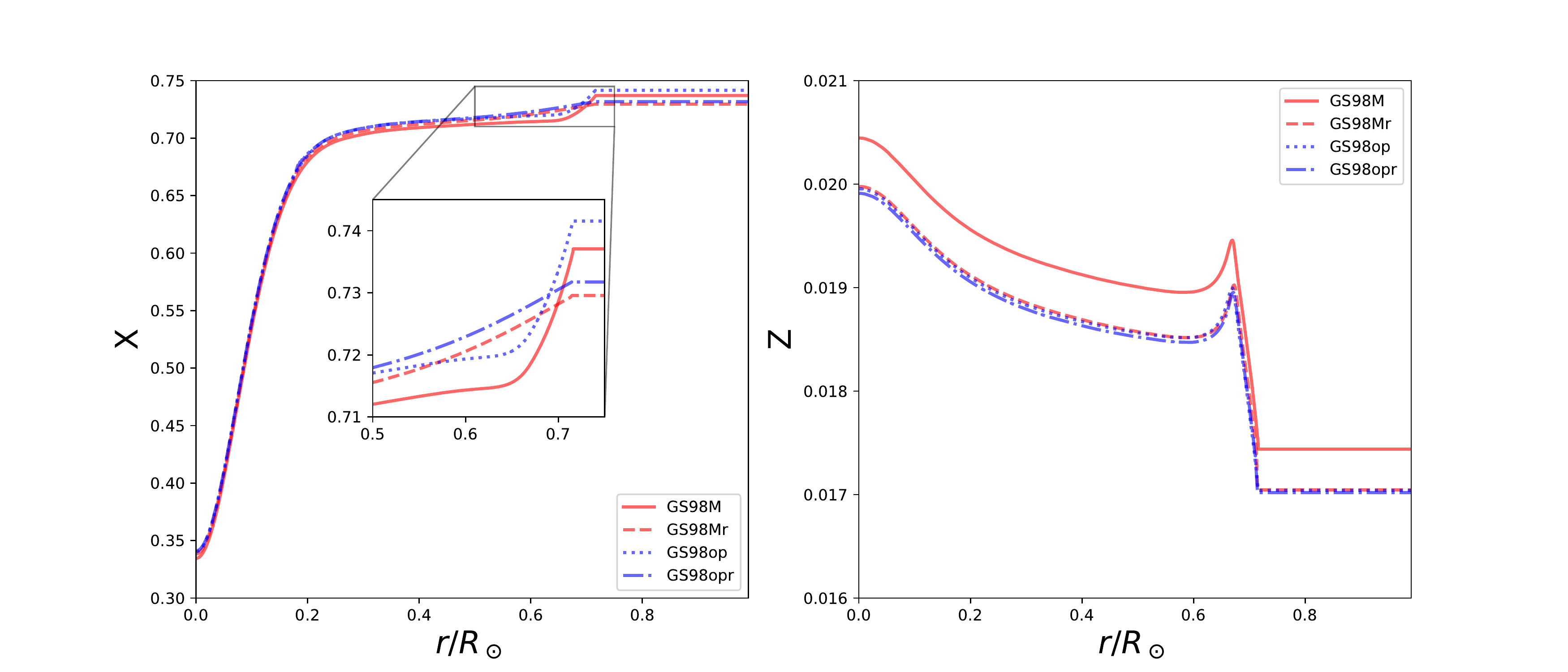}
\caption{Distributions of hydrogen and metal abundances of models as a function of radius.
\label{fig4}}
\end{figure*}

Rotational mixing brings hydrogen into inner layers from outer layers and transports helium outward,
i.e. decreases the hydrogen abundance in the CZ, but increases the hydrogen abundance in the region
with $0.5$ \dsr{} $\lesssim r \lesssim 0.7$ \dsr{} (see Figure \ref{fig4}), which changes the
distribution of the mean molecular weight. As a consequence, the density and sound-speed profiles
are significantly improved by the effects of rotation. The value of the relative sound-speed
difference below the CZ is decreased by about $80\%$ (see Figure \ref{fig1}). Although OP
improves the sound-speed profile below the CZ, the effects of rotation play a more important
role in improving the sound-speed profile. Moreover, the amount of the CZ helium settling
is reduced by about $34\%$. The surface helium abundance of $0.2511$ for GS98opr is in agreement
with the seismically inferred value of $0.2485\pm0.0035$, and increased by about $0.01$ compared
to that of the non-rotating model.

The central temperatures of models constructed with OP opacity tables are lower than those of
models constructed with OPAL opacity tables. Therefore, the fluxes of $^{7}$Be, $^{8}$B,
$^{13}$N, $^{15}$O, and $^{17}$F neutrinos computed from GS98op and GS98opr are smaller than
those calculated from GS98M and GS98Mr. The total fluxes $\Phi(\rm CNO)$ of GS98op and GS98opr
are in agreement with that detected by \citet{bore20}. The $^{8}$B neutrino fluxes of the models
are consistent with that determined by \citet{berg16} but lower than one detected by \citet{bore18}.
Thus the high-Z models constructed with OP opacities also do not completely agree with helioseismic
results and updated neutrino fluxes.

We chose GS98M as the best SSM and GS98opr as the best rotating model with high metal abundances
and went on to construct the models with low metal abundances and compare them with
these two models.

\subsection{Solar Models with Low Metal Abundances }
\subsubsection{Standard and Rotating Models}

\begin{figure*}
\includegraphics[angle=-90, scale=0.7]{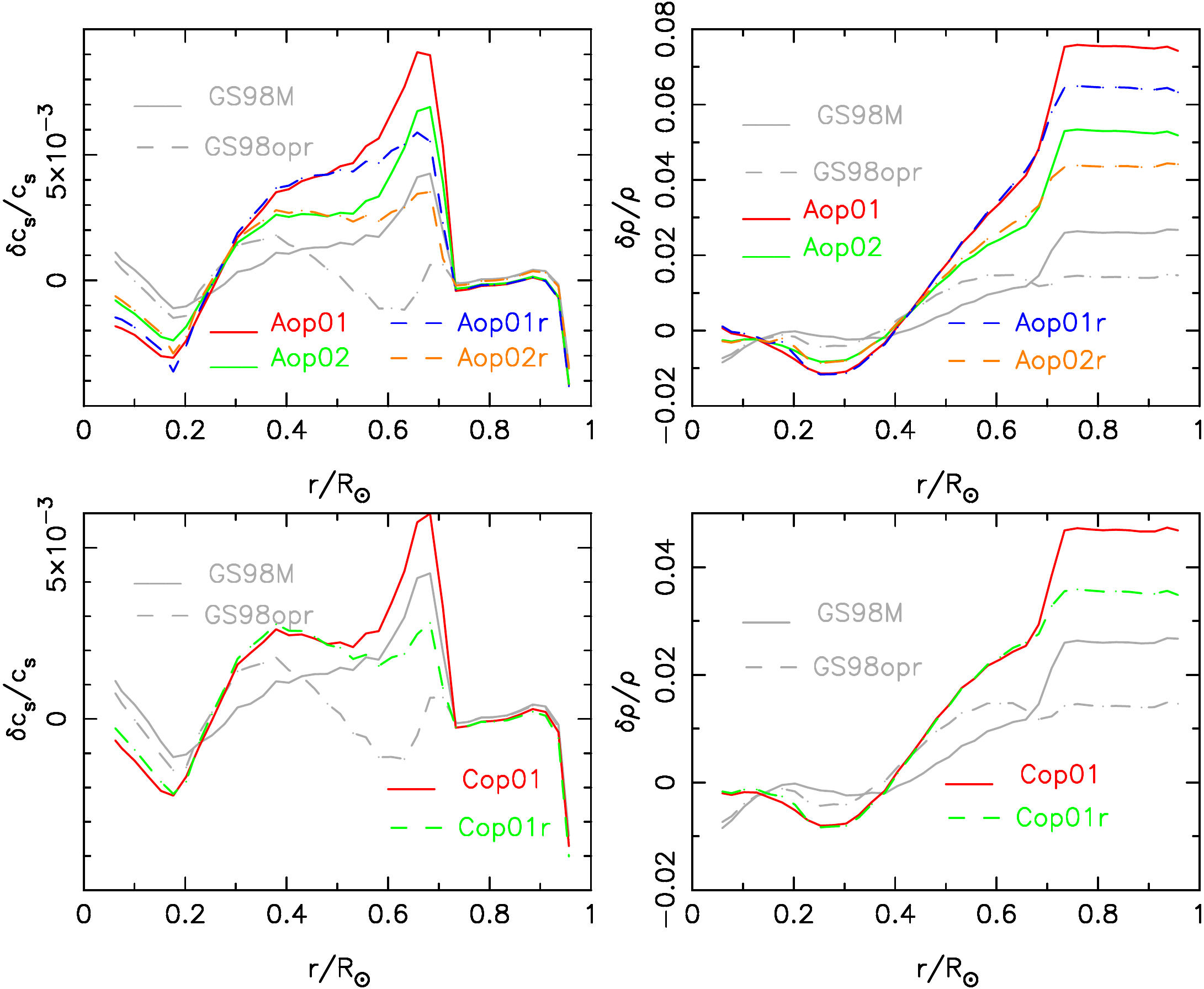}
\caption{Relative sound-speed and density differences, in the sense (Sun-Model)/Model,
between the Sun and models. The solar sound speed and density are given in \citet{basu09}.
\label{fig5}}
\end{figure*}

\begin{figure*}
\includegraphics[angle=-90, scale=0.8]{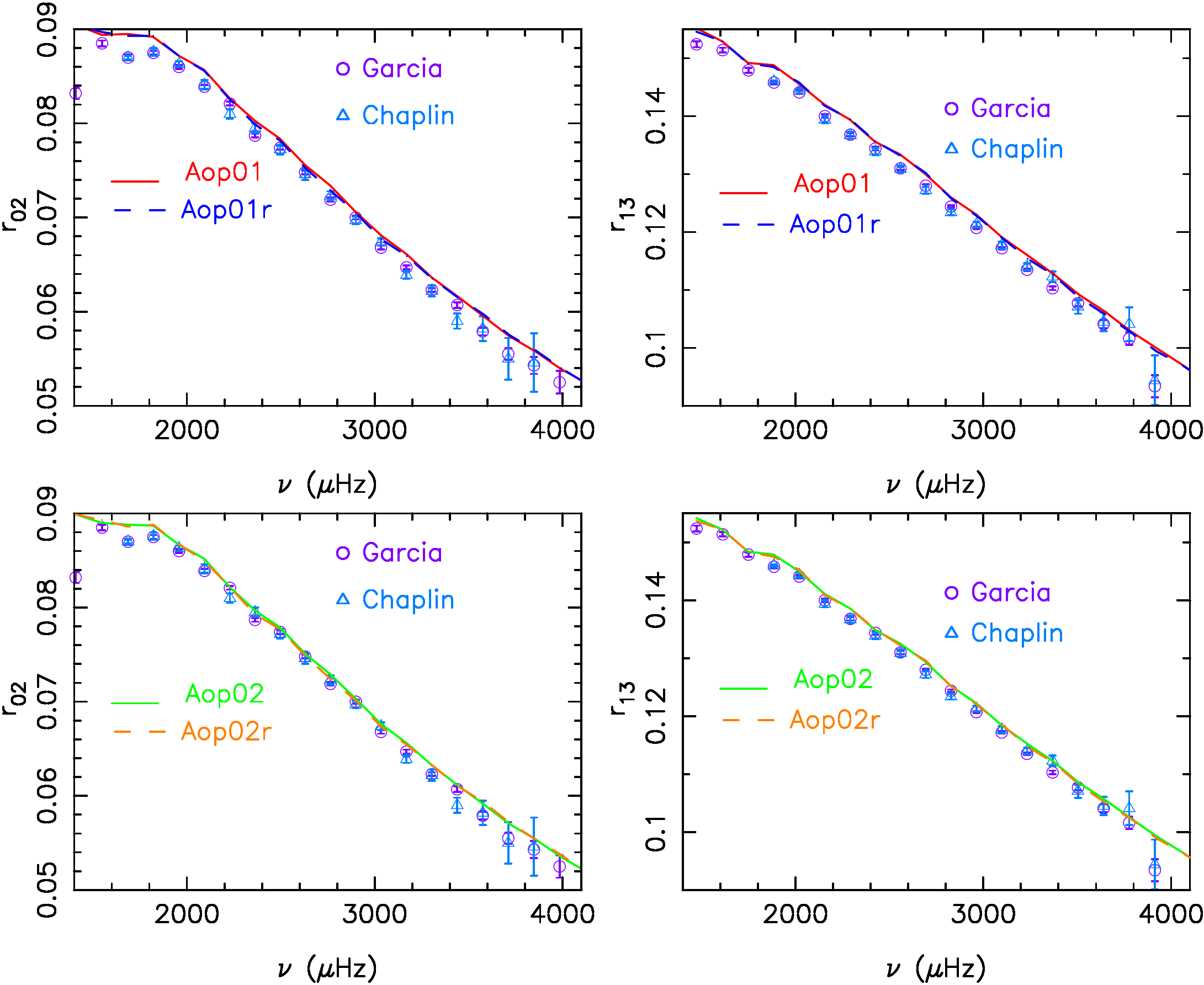}
\includegraphics[angle=-90, scale=0.8]{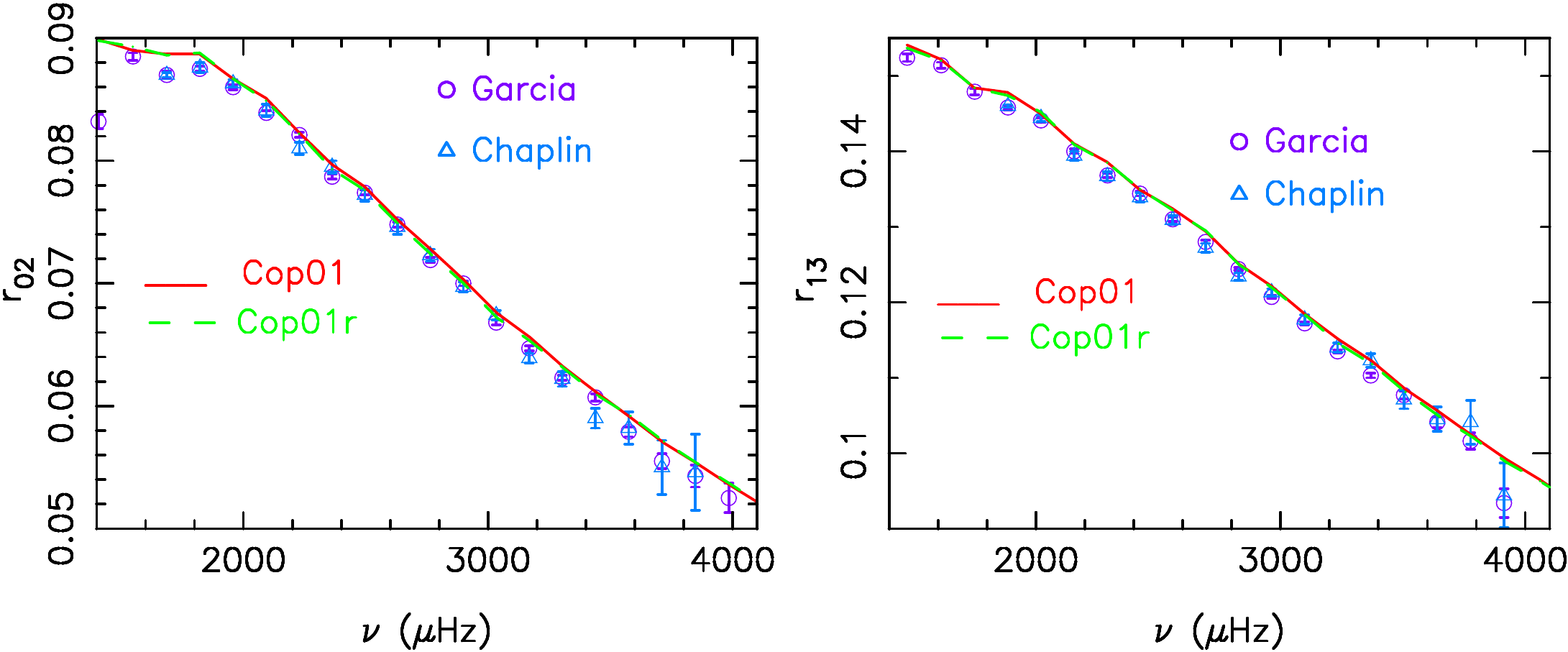}
\caption{Distributions of observed and predicted ratios $r_{02}$ and $r_{13}$ as a function
of frequency. The circles and triangles show the ratios calculated from the frequencies observed
by GOLF \& VIRGO \citep{gar11} and BiSON \citep{cha99b}, respectively.
\label{fig6}}
\end{figure*}

Using OP opacity tables reconstructed with AAG21 mixtures, we computed SSM Aop01 and
rotating model Aop01r with the surface heavy-element abundance determined by \citet{aspl21}.
Although OP opacities and the effects of rotation can significantly improve
sound-speed and density profiles, the sound-speed and density profiles of Aop01 and Aop01r
are not as good as those of GS98M (see Figure \ref{fig5}). The models also can not reproduce
the observed frequency separation ratios $r_{02}$ and $r_{13}$ (see Figure \ref{fig6} or
Table \ref{tab4}) and inferred helium abundance. The total fluxes $\Phi(\rm CNO)$ of Aop01 and Aop01r
are in agreement with the detection of \citet{bore20}, but their $^{7}$Be and $^{8}$B neutrino fluxes
are too low (see Table \ref{tab2}).

With the heavy-element abundance determined by \citet{lod20}, we constructed
SSM Aop02 and rotating model Aop02r. We also calculated SSM Cop01 and rotating model
Cop01r by using the OP opacity tables reconstructed with \citet{caf11} mixtures.
The SSMs Aop02 and Cop01 obviously disagree with seismically inferred results and
detected $^{8}$B neutrino flux. For rotating models Aop02r and Cop01r, the overshoot
of convection brings the depth of the CZ into agreement with the seismically inferred one.
The surface helium abundances of Aop02r and Cop01r are $0.2453$ and $0.2467$, respectively,
consistent with the inferred value of $0.2485\pm0.0035$. The total fluxes $\Phi(\rm CNO)$
predicted by Aop02r and Cop01r are $7.4$ and $7.6\times10^{8}$ cm$^{-2}$ s$^{-1}$, respectively,
which are in good agreement with that detected by \citet{bore20}. The sound-speed profiles
of Aop02r and Cop01r are comparable with that of GS98M (see Figure \ref{fig5}).
The observed frequency separation ratios $r_{02}$ and $r_{13}$ are almost reproduced by
the models (see Figure \ref{fig6}). However, the $^{8}$B neutrino fluxes of Aop02r and Cop01r
are obviously lower than those determined by \citet{berg16} and \citet{bore18} (see Table \ref{tab2}).
Moreover, their density profiles are not as good as that of GS98M. These indicate that the effects
of OP and rotation can improve the solar model but can not completely solve the solar modeling problem.

\subsubsection{Rotating and Enhanced Diffusion Models Constructed in Accordance with AAG21 Mixtures}
The gravitational settling and diffusion reduce the surface helium abundance of SSMs by
around $11\%$ ($\sim0.03$ by mass fraction) below its initial value (see Table \ref{tab1}),
which plays an important role in shaping the sound-speed and density profiles of the models.
Rotational mixing reduces the amount of the surface helium settling by about $33\%-36\%$.
In order to counteract the effects of rotational mixing on the settling of elements, we
increased the rates of element diffusion and settling by $36\%$, and then constructed
rotating models Aop12r and Aopal12r by using OP and OPAL opacity tables. The two models
had the surface heavy-element abundance determined by \citet{lod20}. Tables \ref{tab1}
and \ref{tab2} list their fundamental parameters and neutrino fluxes, respectively.

\begin{figure*}
\includegraphics[angle=-90, scale=0.8]{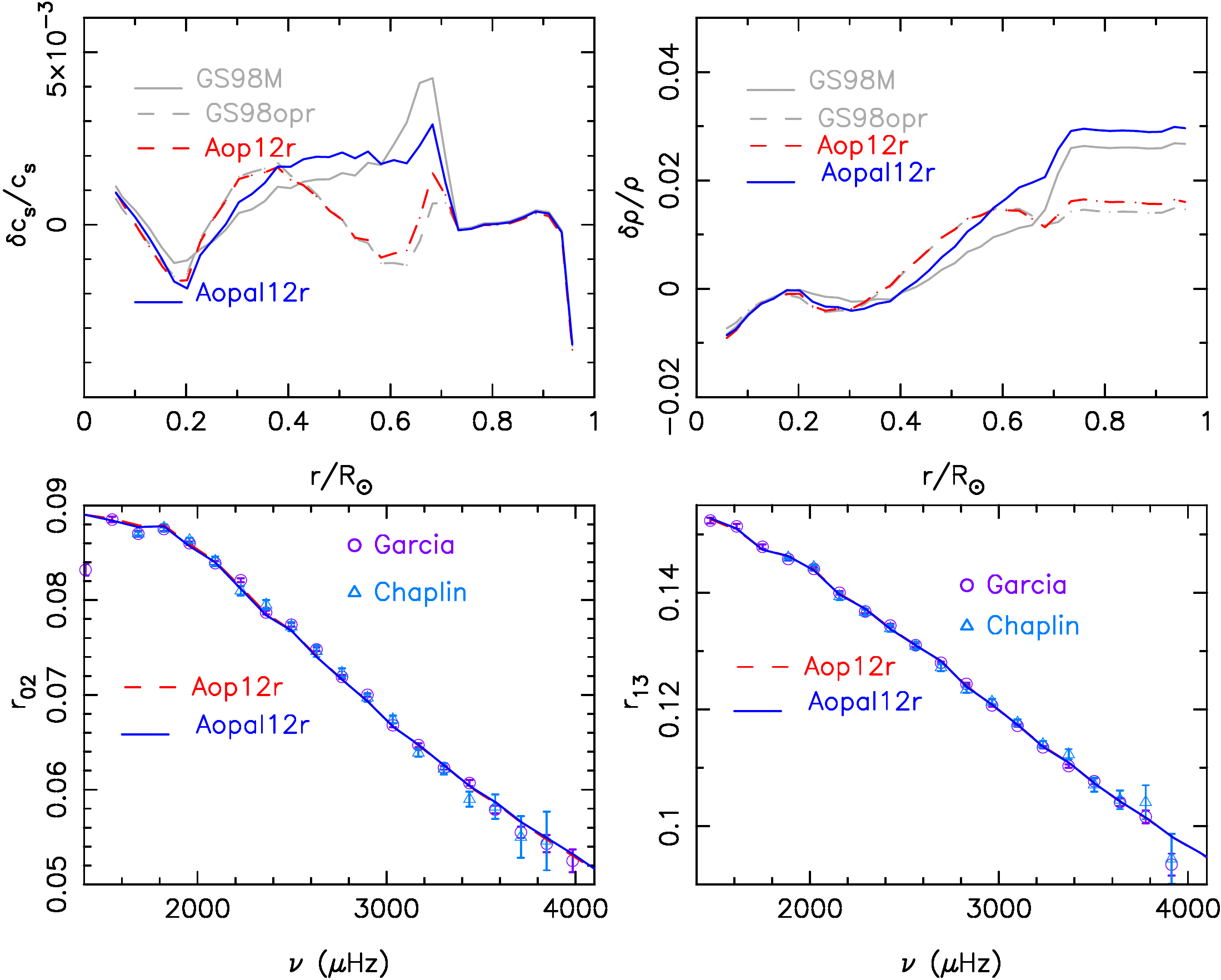}
\caption{Top panels: relative sound-speed and density differences, in the sense (Sun-Model)/Model,
between the Sun and models. The solar sound speed and density are given in \citet{basu09}.
Bottom panels: distributions of observed and predicted ratios $r_{02}$ and $r_{13}$ as a function
of frequency. The circles and triangles show the ratios calculated from the frequencies observed
by GOLF \& VIRGO \citep{gar11} and BiSON \citep{cha99b}, respectively.
\label{fig7}}
\end{figure*}

The sound-speed and density profiles of Aopal12r are almost as good as those of GS98M,
but those of Aop12r are obviously better than those of GS98M (see Figure \ref{fig7}).
The inferred CZ depth and observed frequency separation ratios $r_{02}$ and $r_{13}$ are reproduced
well by the two models (see Table \ref{tab1} and Figure \ref{fig7}). The total fluxes of
$^{13}$N, $^{15}$O, and $^{17}$F neutrinos are $\Phi(\rm CNO)=8.37\times10^{8}$ cm$^{-2}$ s$^{-1}$
for Aop12r and $\Phi(\rm CNO)=8.53\times10^{8}$ cm$^{-2}$ s$^{-1}$ for Aopal12r, which are
consistent with the detected value of $7^{+3}_{-2}\times10^{8}$ cm$^{-2}$ s$^{-1}$ \citep{bore20}.
However, the $^{8}$B neutrino fluxes computed from Aop12r and Aopal12r are lower than that
detected by \citet{bore18} (see Table \ref{tab2}). The surface helium abundances of Aop12r
and Aopal12r are $0.2413$ and $0.2424$, respectively, which are consistent with $0.2423\pm0.0054$
advocated by \citet{aspl21} but lower than $0.2485\pm0.0035$ inferred by \citet{basu04}.
Thus these models also do not completely agree with helioseismic results and updated neutrino fluxes.

\subsubsection{Rotating and Enhanced Diffusion Models Constructed in Accordance with Caffau Mixtures}

\begin{figure*}
\includegraphics[angle=-90, scale=0.8]{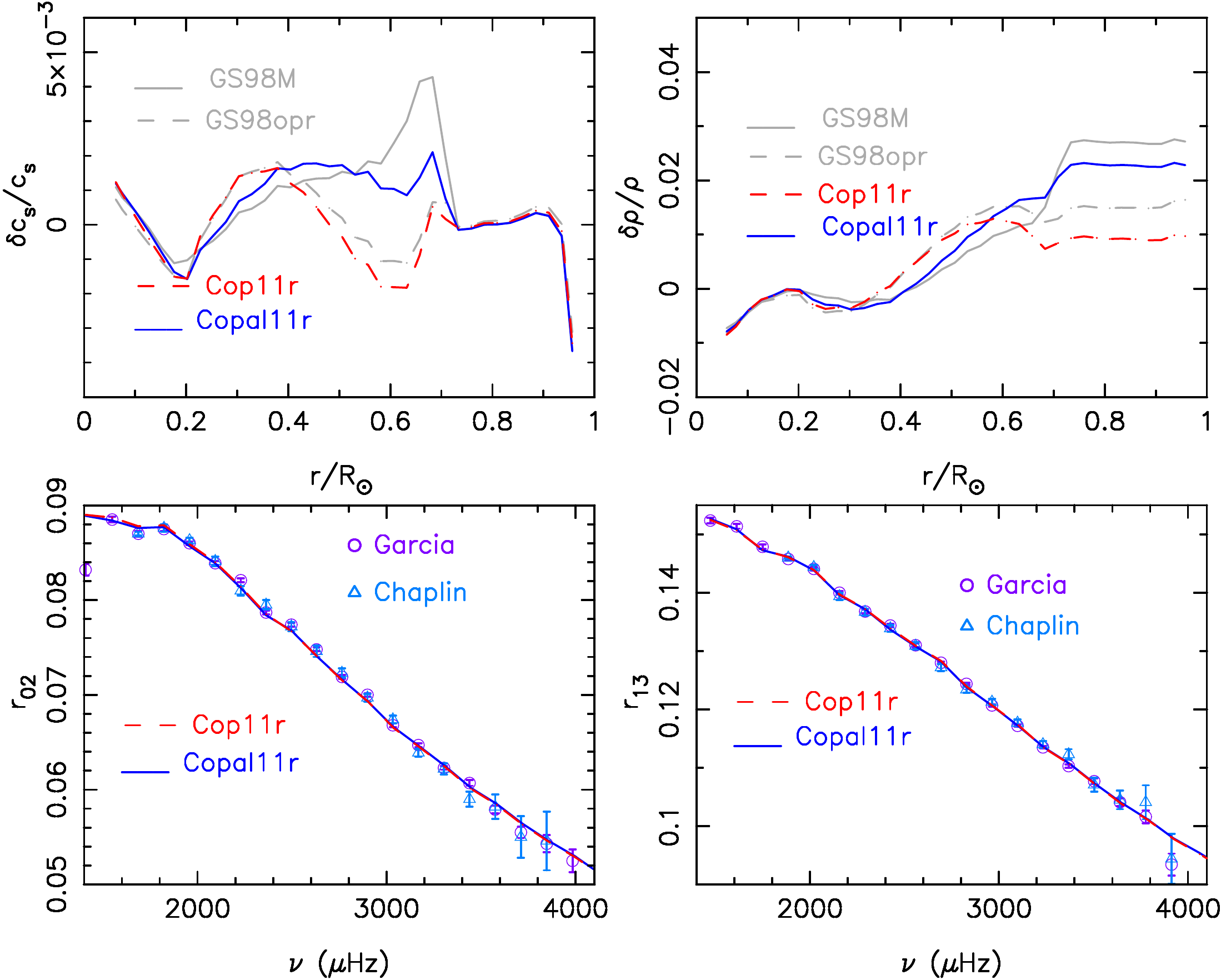}
\caption{Top panels: relative sound-speed and density differences, in the sense (Sun-Model)/Model,
between the Sun and models. The solar sound speed and density are given in \citet{basu09}.
Bottom panels: distributions of observed and predicted ratios $r_{02}$ and $r_{13}$ as a function
of frequency. The circles and triangles show the ratios calculated from the frequencies observed
by GOLF \& VIRGO \citep{gar11} and BiSON \citep{cha99b}, respectively.
\label{fig8}}
\end{figure*}

By using OPAL and OP opacity tables reconstructed with \citet{caf11} mixtures, we computed rotating
models Copal11r and Cop11r. In order to counteract the effects of rotational mixing on the
settling of elements, same as the cases in models Aopal12r and Aop12r, the rates of element diffusion
and settling were also increased by $36\%$ in models Copal11r and Cop11r. The surface
heavy-element abundance of $0.01548$ for Copal11r is consistent with that determined
by \citet{caf10}. Figure \ref{fig8} shows that Copal11r has better sound-speed and density profiles
(smaller $\chi^{2}_{c_{s}+\rho}$) than GS98M and reproduces the observed frequency separation
ratios $r_{02}$ and $r_{13}$. The relative differences $\delta c_{s}/c_{s}$ and $\delta \rho/\rho$
between the Sun and Copal11r are smaller than $0.0021$ and $0.025$, respectively. It also reproduces
the seismically inferred surface helium abundance and radius $r_{cz}$ at the level of $1\sigma$ (see Table \ref{tab1}).

The fluxes of $pp$, $pep$, $hep$, $^{7}$Be, and $^{8}$B neutrinos and the total fluxes of $^{13}$N,
$^{15}$O, and $^{17}$F neutrinos calculated from Copal11r are in agreement with those detected
by \citet{bore18, bore20} at the level of $1\sigma$. The $^{8}$B neutrino flux of $5.41 \times10^{6}$
cm$^{-2}$ s$^{-1}$ is also in good agreement with $(5.21\pm 0.27)\times10^{6}$ cm$^{-2}$ s$^{-1}$
\citep{ahm04} but larger than that determined by \citet{berg16} (see Table \ref{tab2}). Copal11r not only
is in agreement with updated neutrino fluxes but has better sound-speed and density profiles than GS98M
(see Figure \ref{fig8} or Table \ref{tab4}). It is thus better than GS98M.

The surface heavy-element abundance of Cop11r is also $0.01548$. Cop11r has better sound-speed and
density profiles (smaller $\chi^{2}_{c_{s}+\rho}$) than Copal11r (see Figure \ref{fig8}
and Table \ref{tab4}). It also reproduces the observed $r_{02}$ and $r_{13}$, inferred radius $r_{cz}$,
and updated neutrino fluxes except the $^{8}$B neutrino flux. The $^{8}$B neutrino flux of Cop11r
is $5.16 \times10^{6}$ cm$^{-2}$ s$^{-1}$, which is lower than $5.68^{+0.39}_{-0.41} \times10^{6}$
cm$^{-2}$ s$^{-1}$ \citep{bore18} but in good agreement with that determined by \citet{berg16}.
The surface helium abundance of $0.2431$ for Cop11r is lower than $0.2485\pm0.0035$ inferred by
\citet{basu04} but consistent with $0.2423\pm0.0054$ advocated by \citet{aspl21}. OP significantly
improves the sound-speed and density profiles, but leads to the fact that the $^{8}$B neutrino flux
and surface helium abundance are lower than the inferred values. However,
if $\Phi(^{8}\rm B)=5.16^{+0.13}_{-0.09} \times10^{6}$ cm$^{-2}$ s$^{-1}$ \citep{berg16}
and $Y_{s}=0.2423\pm0.0054$ \citep{aspl21} are adopted, the $\Phi(^{8}\rm B)$ and $Y_{s}$ of GS98op
are more consistent with these values than those of GS98M. In this case, GS98op is the best SSM
with high metal abundances rather than GS98M; and Cop11r is better than GS98op.

Models Cop11r and Copal11r have the same input physics except opacity tables.
The models constructed with OP opacity tables have better sound-speed and density profiles
but lower surface helium abundance and $^{8}$B neutrino flux than those constructed with
OPAL opacity tables. The differences in the sound-speed and density profiles, surface helium
abundances, and $^{8}$B neutrino fluxes caused by the discrepancies in the opacities
are marked. The results indicate that a small discrepancy in opacities can obviously affect
the solar model.

\subsection{Possible Problems in Opacities}
\label{inopacity}
\begin{figure*}
\includegraphics[angle=0, scale=0.5]{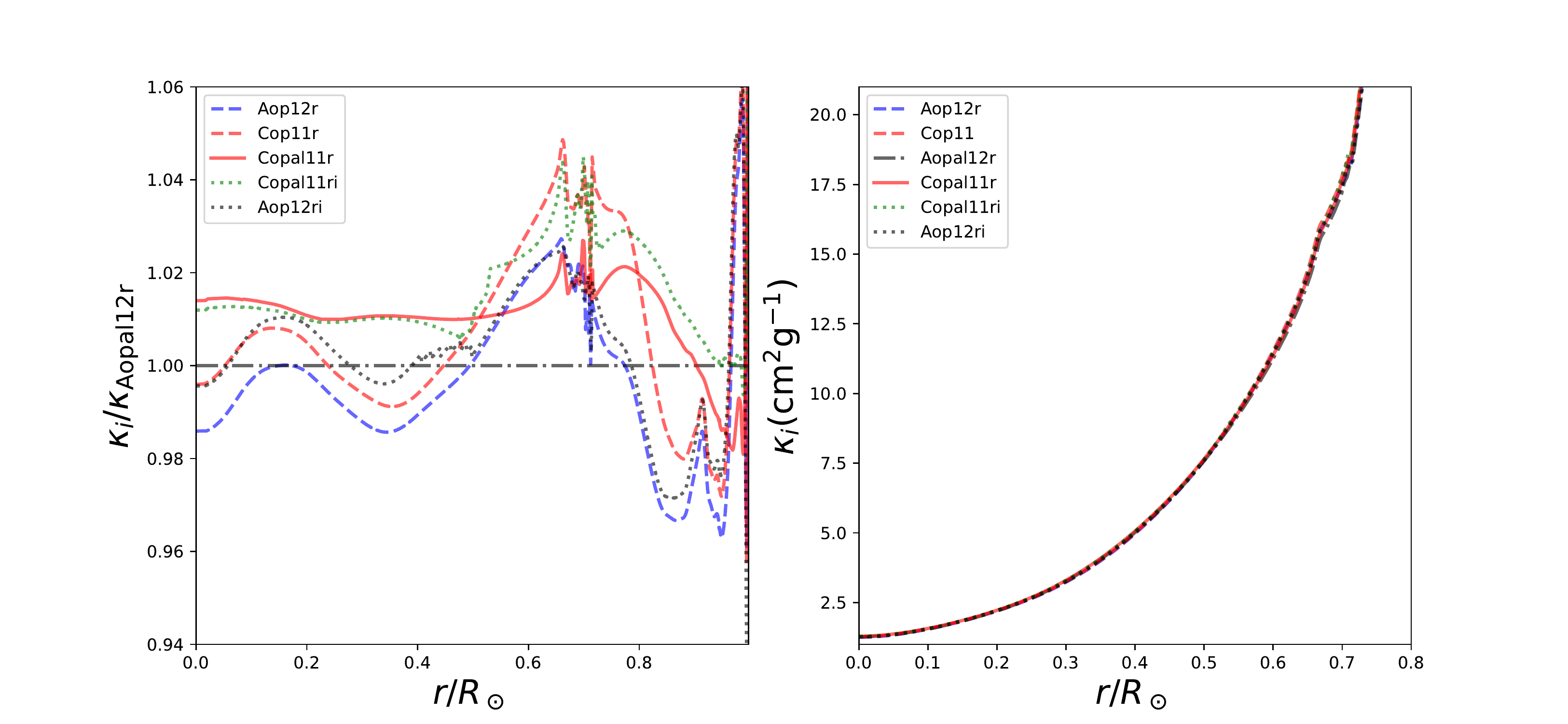}
\caption{Left panel: comparison of Rosseland mean opacity of different models relative to that of Aopal12r.
Right panel: distributions of Rosseland mean opacity of the models as a function of radius.
\label{fig9}}
\end{figure*}

In order to understand the effects of the differences in opacities on models, we constructed model
Aop12ri with OP opacity tables and Copal11ri with OPAL opacity tables. The OP opacities in the
regions of Aop12ri with $T\gtrsim 4.5\times 10^{6}$ K ($r\lesssim 0.45$ \dsr{}) were increased by
about $1.5\%$ ($k_{0}\leq1.015$, see Appendix \ref{appda}) to approximately match OPAL opacities
in the same regions. The OPAL opacities for the regions of Copal11ri with $2\times 10^{6}$ K
$\lesssim T \lesssim 5\times 10^{6}$ K were increased by about $1-2.3\%$ ($k_{0}\leq1.023$)
to roughly match OP opacities (see Figure \ref{fig9}). The fundamental parameters and neutrino fluxes of
these two models are also listed in Tables \ref{tab1} and \ref{tab2}, respectively.

\begin{figure*}
\includegraphics[angle=-90, scale=0.8]{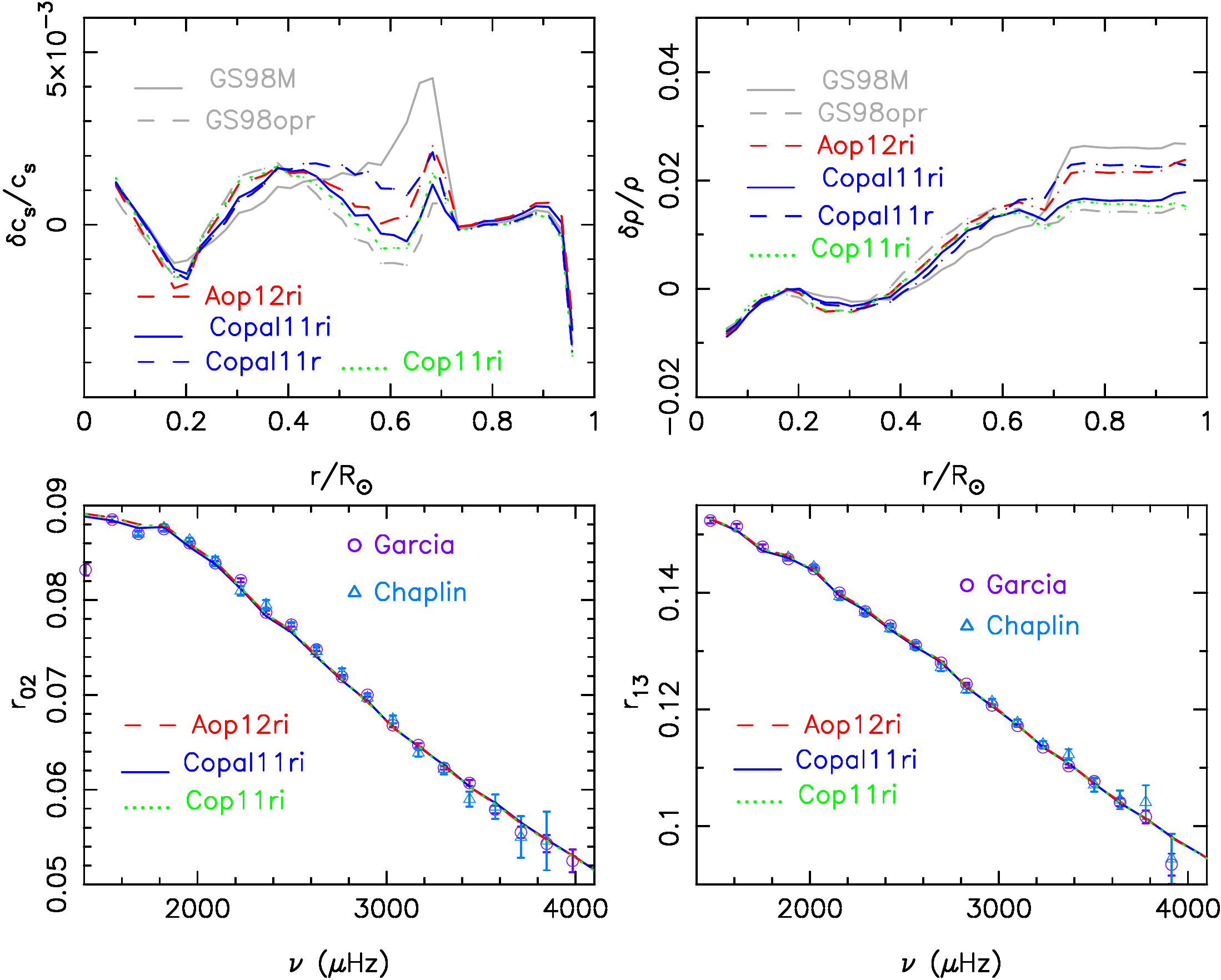}
\caption{Top panels: relative sound-speed and density differences, in the sense (Sun-Model)/Model,
between the Sun and models. The solar sound speed and density are given in \citet{basu09}.
Bottom panels: distributions of observed and predicted ratios $r_{02}$ and $r_{13}$ as a function
of frequency. The circles and triangles show the ratios calculated from the frequencies observed
by GOLF \& VIRGO \citep{gar11} and BiSON \citep{cha99b}, respectively.
\label{fig10}}
\end{figure*}

Aop12ri is in good agreement with helioseismic results and updated neutrino fluxes except the surface
helium abundance and $^{8}$B neutrino flux. The surface helium abundance of Aop12ri is $0.2436$, which
is lower than the seismically inferred value of $0.2485\pm0.035$. Moreover, the $^{8}$B neutrino
flux of Aop12ri is $5.19 \times10^{6}$ cm$^{-2}$ s$^{-1}$, which is lower than that detected by \citet{bore18}.
But they are higher than those of Aop12r. The increase in OP opacities mainly improves the predictions
of surface helium abundance and $^{8}$B neutrino flux but slightly worsens the sound-speed and
density profiles in comparison to those of Aop12r (see Table \ref{tab4} or Figures \ref{fig8} and \ref{fig10}).
However, the sound-speed and density profiles of Aop12ri are still better
than those of GS98M.

The increase in OPAL opacities has almost no effect on surface helium abundance and neutrino fluxes
(see Table \ref{tab2}) but significantly improves the sound-speed and density profiles (see
Figure \ref{fig10}). The relative differences $\delta c_{s}/c_{s}$ and $\delta \rho/\rho$ between
Sun and Copal11ri are smaller than $0.00117$ and $0.0178$ near the BCZ, decreased by about $46\%$
and $29\%$ in comparison to those of Copal11r, respectively. Copal11ri reproduces the observed
ratios $r_{02}$ and $r_{13}$ (see Figure \ref{fig10}) and the inferred helium abundance and $r_{cz}$
at the level of $1\sigma$. The neutrino fluxes calculated from Copal11ri also agree with those detected
by \citet{bore18, bore20} at the level of $1\sigma$. Copal11ri is the best rotating model for
the heavy-element abundance determined by \citet{caf10}. It is better than Copal11r and in good
agreement with helioseismic results and updated neutrino fluxes.

The increase in OPAL opacities works well. The changes in OP opacities would have the same effect.
In order to test this case, we constructed model Cop11ri. The OP opacities for the regions of Cop11ri with
$T\gtrsim 4.5\times 10^{6}$ K were increased by about $1.5\%$ to approximately match OPAL opacities
in the same regions. The calculations show that models Cop11ri and Copal11ri have almost the same sound-speed
and density profiles (see Figure \ref{fig10}), surface helium and heavy-element abundances (see Table \ref{tab1}),
and neutrino fluxes (see Table \ref{tab2}). They also have almost the same $Z_{0}$, $Y_{0}$, and \dalpha{}.
The modified OP and OPAL opacities produce almost the same rotating models. These imply that the differences
between Cop11r and Copal11r result from discrepancies in opacities, and that OPAL might
underestimate opacities for the regions of the Sun with $2\times 10^{6}$ K
$\lesssim T\lesssim 5\times 10^{6}$ K by about $1-2\%$, or that OP might underestimate opacities
in the regions of the Sun with $T \gtrsim 4.5\times 10^{6}$ K by around $1.5\%$.
The possible underestimate in OP leads to the fact that the models constructed with
OP opacity tables have a lower surface helium abundance and $^{8}$B neutrino flux than those
constructed with OPAL opacity tables, while the possible underestimate in OPAL results in
the fact that the sound-speed and density profiles of the models constructed with OPAL opacity tables
are not as good as those of the models constructed with OP opacity tables.

With the modified opacities, only the models with the metal abundance determined by \citet{lod20}
agree with the $^{8}$B neutrino flux determined by \citet{berg16} and the helium abundance advocated
by \citet{aspl21}, but the models with the metal abundance determined by \citet{caf10} are in
agreement with the $^{8}$B neutrino flux detected by \cite{bore18} and the seismically inferred
helium abundance. Thus precisely determining $^{8}$B neutrino flux aids in solving
the solar abundance problem.

\section{Discussion and Summary}

Although GS98M was chosen as the best SSM with high metal abundances, the value of $\Phi(\rm CNO)$ of GS98M
is slightly larger than the detected value of $7^{+3}_{-2}\times10^{8}$ cm$^{-2}$ s$^{-1}$ \citep{bore20}.
That of GS98opr is in agreement with the detected one, but its $^{8}$B neutrino flux is lower
than that updated by \citet{bore18}. Thus the updated neutrino fluxes do not favour the high-Z models.
Moreover, the values of $\Phi(\rm CNO)$ calculated from the models with the heavy-element abundance determined by
\citet{aspl21} are consistent with the detected value, but the $^{7}$Be and $^{8}$B neutrino fluxes
predicted by the models are much lower than those determined by \citet{berg16} and ones
detected by \citet{bore18}. Therefore, the updated neutrino fluxes also do not prefer the models with
the heavy-element abundance determined by \citet{aspl21}. For the same input physics, the fluxes of $^{7}$Be,
$^{8}$B, $^{13}$N, $^{15}$O, and $^{17}$F neutrinos predicted by models increase with an increase in metallicity.
These imply that the updated neutrino fluxes prefer a heavy-element abundance between that determined by GS98
and one advocated by \citet{aspl21}.

Convection was treated according to the standard mixing-length theory \citep{boh58} in this work.
The treatment of convection is one of the sources of uncertainty in modeling of stars.
Different treatments of convection, such as \citet{joyc18}, \citet{spad18, spad19},
and \citet{jerm22}, could affect solar models and deserve more detailed study.

There are many $T-\tau$ relations for solar atmosphere \citep{kris66, ball22}. The temperature, $T$ reaches
\dteff{} at $\tau=2/3$ for the Eddington approximation but $\tau=0.312156330$ for the \citet{kris66} $T-\tau$
relation, i.e. the solar radius R is defined at an outer layer for the \citet{kris66} $T-\tau$
relation. As a consequence, the \citet{kris66} $T-\tau$ relation requires a larger \dalpha{} to
reproduce solar radius than the Eddington approximation \citep{dema08, joyc18, spad18}. In order to
study the effect of the $T-\tau$ relation on our results, we computed the solar models with
the Eddington approximation. The calculations show that the $T-\tau$ relation has hardly
any influence on neutrino fluxes, frequency separation ratios, and sound-speed and density profiles
in the radiative region, but slightly affects the sound-speed and density profiles in the CZ (lead to
a slight increase in $\chi^{2}_{c_{s}+\rho}$). Choosing between the Eddington approximation or
\citet{kris66} $T-\tau$ relation does not change our results.

Rotational mixing can more efficiently inhibit the settling of helium than of heavy-element
abundances because the mixing depends on the gradient of elements \citep{yang19}. It can reduce
the amount of the surface helium settling by about $33-36\%$ in our models, which is consistent with
the result of \citet{pro91}, who found that macroscopic turbulent mixing can reduce the amount of
the surface helium settling by around $40\%$. It leads to the fact that the surface helium abundances
of rotating models are obviously higher than those of non-rotating models. In the enhanced diffusion models,
the velocity of diffusion and settling was increased by $36\%$. However, we have no obvious physical
justification for the multiplier. In non-rotating models, the enhanced diffusion leaves the surface
helium abundance too low. However, rotational mixing completely counteracts the effect of the enhanced
diffusion on the surface helium abundance in rotating models. Thus the surface helium abundances of
the rotating models with the enhanced diffusion are higher than those of non-rotating models.
The effects of rotation and enhanced diffusion bring low-Z models into agreement with helioseismic
results \citep{basu04, basu09} and updated neutrino fluxes \citep{bore18, bore20}. However, the calculations
show that the same effects can not bring high-Z models into agreement with the helioseismic results and
the updated neutrino fluxes at the same time.

The effects of rotation and enhanced diffusion were studied by \citet{yang19}, where the value of
multiplier ($f_{0}$) was larger than or equal to $1.5$ and OPAL opacity tables constructed in accordance
with AGSS09 mixtures were used. The flux of $^{7}$Be neutrino and the total fluxes of $^{13}$N, $^{15}$O,
and $^{17}$F neutrinos predicted by the best model of \citet{yang19} are larger than those detected
by \citet{bore18, bore20}. The $^{7}$Be neutrino flux is also higher than that determined by \citet{berg16}.
Different from the earlier models of \citet{yang19}, Copal11r and Copal11ri, in which the value of
multiplier is $1.36$ and opacity tables are reconstructed in accordance with \citet{caf11} mixtures,
are in good agreement with the detected neutrino fluxes at the level of $1\sigma$.

The differences between OPAL and OP opacities are small but can obviously affect the properties of
solar models. The $^{8}$B neutrino fluxes computed from models constructed with OP opacity tables
are lower than those calculated from models constructed with OPAL opacity tables, which derives
from the fact that OP underestimates opacities for the regions of the Sun with $T \gtrsim 5\times 10^{6}$
K by about $1.5\%$ compared to OPAL, especially in the core. As a consequence, the models constructed
with OP opacity tables disagree with the $^{8}$B neutrino flux detected by \citet{bore18} but can agree
with that determined by \citet{berg16}. Thus precisely determining the $^{8}$B neutrino flux aid in
diagnosing the opacities in the solar core.

Moreover, the sound-speed and density profiles of models constructed with OPAL opacity tables are obviously
not as good as those of models constructed with OP opacity tables, which results from the fact that OPAL
underestimates opacities for the regions of the Sun with $2\times 10^{6}$ K $\lesssim T\lesssim 5\times 10^{6}$
K by about $1-2\%$ compared to OP. If OPAL opacities in the regions are increased by about $1-2\%$ to
approximately match OP opacities for the same regions, the models constructed with the OPAL
opacities will produce better sound-speed and density profiles. If OP opacities for the regions
of the Sun with $T \gtrsim 5\times 10^{6}$ K are increased by about $1.5\%$, the model constructed with
the OP opacities will be almost the same as that constructed with the modified OPAL opacities.
These imply that the discrepancies between the sound speed and density of the Sun and those of
the models could partly derive from opacity. The small differences between OPAL and OP
opacities can obviously affect sound-speed and density profiles, surface helium abundance, and neutrino
fluxes of models, which do not depend on mixture patterns. But the effect on neutrino fluxes
slightly relies on the value of $Z_{0}$. In order to improve the solar model, the discrepancies between
OPAL and OP may deserve more studies.

In this work, by using OP and OPAL opacity tables reconstructed with AAG21 and Caffau's mixtures,
we constructed rotating solar models in which the effects of convection overshoot and enhanced
diffusion were included. We obtained a rotating model, Copal11r, that is better than the SSM GS98M
and the earlier rotating models of \citet{yang16, yang19}. The surface heavy-element abundance of
Copal11r is $0.01548$, which is consistent with the value of $0.0154$ determined by \citet{caf10}
and that inferred by \citet{basu04}. The surface helium abundance of $0.2450$ and the radius of
the BCZ of $0.714$ \dsr{} are in agreement with the seismically inferred values at the level of
$1\sigma$. The initial helium abundance is $0.2718$, which is consistent with the value of $0.273\pm0.006$
inferred by \citet{sere10}. The ratios $r_{02}$ and $r_{13}$ of Copal11r agree with
those calculated from observed frequencies. The sound-speed and density profiles of Copal11r
are better than those of GS98M. Moreover, the fluxes of $pp$, $pep$, $hep$, $^{7}$Be, and $^{8}$B
neutrinos and the total fluxes of $^{13}$N, $^{15}$O, and $^{17}$F neutrinos calculated from Copal11r
agree with those detected by \citet{bore18, bore20} at the level of $1\sigma$. The fluxes of $^{7}$Be
and $^{8}$B neutrinos are also consistent with those determined by \citet{bel11} and \citet{ahm04}.
To recover the seismically inferred depth of the CZ, a convection overshoot of $\delta_{\rm ov}\approx 0.1$
is required in rotating models. Rotation or enhanced diffusion alone can improve sound-speed and
density profiles, but the combination of rotation and enhanced diffusion is required to bring the
rotating model into agreement with seismically inferred results and detected neutrino fluxes.
More details of Copal11r are given in Appendix \ref{appdc}.

If the surface helium abundance of $0.2423\pm 0.0054$ advocated by \citet{aspl21} and the $^{8}$B
neutrino flux determined by \citet{berg16} are adopted, we can obtain another rotating model, Cop11r,
that is better than GS98op. The models constructed with OP opacity tables have obviously better
sound-speed and density profiles but lower surface helium abundance and $^{7}$Be, $^{8}$B, $^{13}$N,
$^{15}$O, and $^{17}$F neutrino fluxes than those constructed with OPAL opacity tables.
The calculations show that OPAL may underestimate opacities for the regions of the Sun with
$2\times 10^{6}$ K $\lesssim T\lesssim 5\times 10^{6}$ K by about $1-2\%$, and that OP may
underestimate opacities in the regions of the Sun with $T \gtrsim 5\times 10^{6}$ K by around $1.5\%$.
If the possible underestimate of OPAL or OP were corrected, the model better than Copal11r or Cop11r
would be obtained.

\begin{acknowledgments}
The author thanks the anonymous referee for helpful comments that helped the author
improve this work, as well as Dr Xianfei Zhang and Tanda Li for their help, and acknowledges
the support from the NSFC 11773005 and U1631236.
\end{acknowledgments}

\clearpage

% \begin{rotatetable}
\begin{deluxetable*}{lllllllllllll}
%\tablenum{1}
\tablecaption{Fundamental Parameters of Solar Models Constructed in Accordance with Different Mixtures.
\label{tab1}}
\tablewidth{0pt}
\tablehead{
 Model &  $Y_{0}$ &  $Z_{0}$ &  \dalpha{} & $\delta_{\rm ov}$ & $f_{0}$ &  $\rho_{c}$ & $r_{cz}$ & $Y_{s}$ & $Z_{s}$ & $(Z/X)_{s}$ &  $\Delta Y$ & $\Omega_{i}$
}
\startdata
\multicolumn{13}{c}{Opacity Tables Constructed with GS98 mixtures}\\
 GS98M  & 0.2761 & 0.01940  & 2.1223 &  0   & 1.0  &  154.59 &  0.716  & 0.2455 & 0.0174 & 0.0237 & 0.0306 & 0  \\
 GS98Mr  & 0.2735 & 0.01896  & 2.0754 & 0.05 & 1.0 &  154.11 &  0.715  & 0.2534 & 0.0170 & 0.0234 & 0.0201 & 10  \\
 GS98op & 0.2713 & 0.01894  & 2.1352 &  0   & 1.0  &  154.25 &  0.716  & 0.2414 & 0.0170 & 0.0230& 0.0299 & 0 \\
 GS98opr& 0.2711 & 0.01890 & 2.0979 & 0.05 & 1.0  &  154.16 &  0.714  & 0.2512 & 0.0170 & 0.0232 & 0.0199 & 10 \\
 \hline
 \hline
 \multicolumn{13}{c}{Opacity Tables Constructed with AAG21 mixtures}\\
 Aop01 & 0.2583  & 0.01557  & 2.0722  &  0   &  1.0  &   151.98  &  0.725  & 0.2283 & 0.0139 & 0.0184 & 0.0300 & 0  \\
 Aop02 & 0.2648  & 0.01661  & 2.0954  &  0   &  1.0  &   152.89  &  0.722  & 0.2347 & 0.0149 & 0.0199 & 0.0301 & 0  \\
 Aop01r & 0.2582 & 0.01554  & 2.0383 & 0.15   &  1.0  &  151.86 &  0.716  & 0.2389 & 0.0140 & 0.0187 & 0.0193 & 10  \\
 Aop02r & 0.2648 & 0.01659  & 2.0596 & 0.15   &  1.0  &  152.88 &  0.713  & 0.2453 & 0.0149 & 0.0202 & 0.0195 & 10  \\
 \hline
 Aop12r  & 0.2675 & 0.01717 & 2.1055 & 0.10   &  1.36 &  154.42  &  0.714 & 0.2413 & 0.0149 & 0.0201 & 0.0262 & 10  \\
 Aopal12r& 0.2690 & 0.01720 & 2.0872 & 0.10 &  1.36  &  154.27  &  0.714 & 0.2424 & 0.0149 & 0.0201 & 0.0266 & 10 \\
 Aop12ri & 0.2702 &0.01722 & 2.0918 & 0.10 &  1.36  &  154.34  &  0.715 & 0.2436 & 0.0149 & 0.0202 & 0.0266 & 10  \\
 \hline
 \hline
 \multicolumn{13}{c} {Opacity Tables Constructed with Caffau mixtures}\\
 Cop01 & 0.2666  & 0.01719  & 2.0924  &  0   &  1.0  &  152.87  &  0.720  & 0.2365 & 0.0154 & 0.0206 & 0.0301 & 0  \\
 Cop01r & 0.2664 & 0.01714 & 2.0582 & 0.12 &  1.0  &  152.76 &  0.714  & 0.2467 & 0.0154 & 0.0209 & 0.0197 & 10 \\
 \hline
 Cop11r & 0.2695 & 0.01780 & 2.1056 & 0.10   &  1.36  &  154.42 &  0.711  & 0.2431 & 0.01548 & 0.0209 & 0.0264 & 10  \\
 Cop11ri & 0.2717 & 0.01777 & 2.0883 & 0.10 & 1.36 &  154.29 & 0.714 & 0.2450 & 0.0154 & 0.0209 & 0.0267 & 10 \\
 Copal11r & 0.2717 & 0.01784 & 2.0817 & 0.10 &  1.36  &   154.25 &  0.713 & 0.2450 & 0.01548 & 0.0209 & 0.0267 & 10  \\
 Copal11ri & 0.2720 & 0.01774 & 2.0865 & 0.10 &  1.36  & 154.46 &  0.712 & 0.2453 & 0.0154 & 0.0209 & 0.0267 & 10 \\
 \enddata
\tablecomments{The central density $\rho_{c}$, CZ radius $r_{cz}$, and initial
angular velocity $\Omega_{i}$ are in units of g cm$^{-3}$, $R_{\odot}$, and $10^{-6}$ rad s$^{-1}$,
respectively. The quantity $\Delta Y=Y_{0}-Y_{s}$ is the amount of surface helium settling.}
\end{deluxetable*}
% \end{rotatetable}

\begin{deluxetable*} {cccccccccc}
%\tablenum{3}
\tablecaption{Predicted and Measured Solar Neutrino Fluxes. \label{tab2}}
% \tablewidth{0pt}
\tablehead{
\colhead{model} & \colhead{$T_{c}$} & \colhead{$pp$} & \colhead{$pep$} &\colhead{$hep$} &\colhead{$^{7}$Be} & \colhead{$^{8}$B} & \colhead{$^{13}$N} & \colhead{$^{15}$O} &\colhead{$^{17}$F}
}
\startdata
 BP04 & ... &  5.94  & 1.40   &  7.88   & 4.86   & 5.79    & 5.71   & 5.03   & 5.91 \\
 SSeM & ... &  5.92  & 1.39   &  ....   & 4.85   & 4.98    & 5.77   & 4.97   & 3.08 \\
 Measured &... & 6.06$^{+0.02 a}_{-0.06}$ & 1.6$\pm$0.3$^{b}$ &...& 4.84$\pm$0.24$^{a}$& 5.21$\pm$0.27$^{c}$ &...& ...&...\\
 B16$^{d}$ &... & 5.97$^{+0.04}_{-0.03}$ & 1.448$\pm$0.013 & 19$^{+12}_{-9}$ & 4.80$^{+0.24}_{-0.22}$ & 5.16$^{+0.13}_{-0.09}$ & $\leq$13.7 & $\leq$2.8 & $\leq$85\\
Borexino$^{e}$ &... & 6.1$\pm$0.5 & 1.39$\pm$0.19 & $<$220 & 4.99$\pm$0.11 & 5.68$^{+0.39}_{-0.41}$ & \multicolumn{2}{c}{7$^{+3}_{-2}$} &  \\
\hline
\hline
% \tableline
 GS98M   & 15.777 & 5.95 & 1.40 &  9.69 & 5.09  & 5.59  & 5.40  & 4.77  & 5.52 \\
 GS98Mr  & 15.733 & 5.96 & 1.41 &  9.73 & 4.95    & 5.29   & 5.08  & 4.45  & 5.14  \\
 GS98op  & 15.706 & 5.98 & 1.42 &  9.81 & 4.92    & 5.18    & 4.97    & 4.35   & 5.02  \\
 GS98opr & 15.693 & 5.97 & 1.41 &  9.78 & 4.86    & 5.07    & 4.90  & 4.28  & 4.93  \\
 \hline
 \hline
 Aop01  & 15.513 & 6.05  & 1.45  &  10.19 & 4.42    & 4.12    & 3.43    & 2.91   & 3.32  \\
 Aop02  & 15.604 & 6.02  & 1.43  &   9.79 & 4.66    & 4.60    & 3.97    & 3.43   & 3.93  \\
 Aop01r & 15.498 & 6.03  & 1.44  &  10.15 & 4.36    & 4.01    & 3.37    & 2.85   & 3.25  \\
 Aop02r & 15.598 & 6.01  & 1.43  &  9.99  & 4.63    & 4.55    & 3.94    & 3.40   & 3.89  \\
 \hline
 Aop12r   & 15.680 & 5.97 & 1.42 & 9.84 & 4.82 & 4.98 & 4.45 & 3.88 & 4.47  \\
 Aopal12r & 15.706 & 5.97 & 1.42 & 9.81 & 4.87 & 5.09 & 4.54 & 3.96 & 4.57 \\
 Aop12ri & 15.715 & 5.96 & 1.41 & 9.77 & 4.91 & 5.19 & 4.61 & 4.04 & 4.66 \\
 \hline
 \hline
 Cop01  & 15.645 & 6.00  & 1.42  &  9.95  & 4.75    & 4.82    & 4.25    & 3.69   & 4.24  \\
 Cop01r & 15.623 & 5.98  & 1.42  &  9.92  & 4.66    & 4.65   & 4.16    & 3.59   & 4.12  \\
 \hline
 Cop11r & 15.716 & 5.95 & 1.41 & 9.77 & 4.89 & 5.16  & 4.73 & 4.15& 4.79\\
 Cop11ri & 15.751 & 5.94 & 1.40 & 9.70 & 4.99 & 5.38 & 4.90 & 4.32 & 4.99 \\
 Copal11r & 15.761 & 5.95 & 1.41  & 9.71 & 4.99 & 5.41 & 4.93  & 4.34 & 5.03  \\
 Copal11ri& 15.764 & 5.95 & 1.41  & 9.72 & 5.00 & 5.44 & 4.92  & 4.34 & 5.02  \\
\enddata
\tablecomments{The table shows the predicted fluxes, in units of $10^{10}(pp)$,
$10^{9}(^{7}\mathrm{Be})$, $10^{8}(pep,^{13}\mathrm{N}, ^{15}\mathrm{O})$,
$10^{6}(^{8}\mathrm{B}, ^{17}\mathrm{F})$, and $10^{3}(hep)$ $\mathrm{cm}^{2}$
$\mathrm{s}^{-1}$. The central temperature $T_{c}$ of models is in unit of $10^{6}$ K.
The BP04 is the best model of \citet{bah04a} and has the GS98 mixtures. The SSeM is
the best seismic model with high metal abundances that can reproduce inferred sound speed \citep{tur11a}.
\tablenotetext{a}{\citet{bel11}.}
\tablenotetext{b}{\citet{bel12}.}
\tablenotetext{c}{\citet{ahm04}.}
\tablenotetext{d}{\citet{berg16}.}
\tablenotetext{e}{The total fluxes produced by CNO cycle are given in \citet{bore20}, other neutrino fluxes
are given in \citet{bore18}.}
}
\end{deluxetable*}

\begin{deluxetable*} {ccccccc}
%\tablenum{3}
\tablecaption{Physical Configurations of Models. \label{tab3}}
% \tablewidth{0pt}
\tablehead{
 Model Name  &  Mixtures  & Opacity & Enhanced Diffusion  & $Z_{s}$ value & Rotation & Increased Opacity ($k_{0}$)$^{a}$
}
\startdata
 GS98M       & GS98 &  OPAL  & No & GS98 & No & No \\
 GS98Mr      & GS98 &  OPAL  & No & GS98 & Yes & No  \\
 GS98op      & GS98 &  OP    & No & GS98 & No & No  \\
 GS98opr     & GS98 &  OP    & No & GS98 & Yes & No  \\
 \hline
 \hline
 Aop01    & AAG21$^{b}$ & OP & No & AAG21         & No & No  \\
 Aop02     & AAG21   &  OP   & No & Lodders 2020  & No & No  \\
 Aop01r    & AAG21   &  OP   & No & AAG21         & Yes & No  \\
 Aop02r    & AAG21   &  OP   & No & Lodders 2020  & Yes & No  \\
 \hline
 Aop12r     & AAG21 &  OP   & Yes & Lodders 2020  & Yes & No  \\
 Aop12ri    & AAG21 &  OP   & Yes & Lodders 2020  & Yes & Yes($1\leq k_{0}\leq1.015$) \\
 Aopal12r   & AAG21 &  OPAL & Yes & Lodders 2020  & Yes & No  \\
 \hline
 \hline
 Cop01     & Caffau$^{c}$ &  OP & No & Caffau    & No  & No  \\
 Cop01r    & Caffau  &  OP  & No  & Caffau       & Yes & No  \\
 \hline
 Cop11r     & Caffau &  OP  & Yes & Caffau     & Yes & No  \\
 Cop11ri    & Caffau &  OP  & Yes & Caffau     & Yes & Yes($1\leq k_{0}\leq1.015$) \\
 Copal11r   & Caffau & OPAL & Yes & Caffau     & Yes & No  \\
 Copal11ri  & Caffau & OPAL & Yes & Caffau     & Yes & Yes($1\leq k_{0}\leq1.023$) \\
\enddata
\tablecomments{$^{a}$ See the Appendix \ref{appda}.
$^{b}$See the Table \ref{tabop} in Appendix \ref{appda} and \citet{aspl21}.
$^{c}$See the Table \ref{tabop} and \citet{caf10, caf11}.
}
\end{deluxetable*}

\begin{deluxetable*} {lcccc}
%\tablenum{3}
\tablecaption{The Values of $\chi^{2}$ of Different Models. \label{tab4}}
% \tablewidth{0pt}
\tablehead{
 Model   &  $\chi^{2\ a}_{c_{s}+\rho}$  & $\chi^{2\ b}_{d_{02+13}}$   & $\chi^{2\ c}_{Borexino-neutrino}$ & $\chi^{2\ d}_{\rm helium}$
}
\startdata
GS98M    & 882.4  & 1.2  & 0.53  & 0.7\\
GS98Mr   & 799.2  & 1.4  & 0.49  & 2.0\\
GS98op   & 578.2  & 1.8  & 0.61  & 4.1\\
GS98opr  & 369.6  & 1.5  & 0.88  & 0.6\\
\hline
\hline
Aop01    &  6244.7 & 19.0 & 7.08  & 33.3\\
Aop02    &  3144.1 & 16.2 & 2.85  & 15.5\\
Aop01r   &  4424.0 & 9.0  & 8.43  & 7.5\\
Aop02r   &  1983.0 & 7.0  & 3.23  & 0.8\\
\hline
Aop12r   &  389.1  & 1.7  & 1.11  & 4.2\\
Aopal12r &  917.4  & 1.6  & 0.78  & 3.0\\
Aop12ri  &  524.3  & 2.0  & 0.57  & 2.0\\
\hline
\hline
Cop01    & 2440.7  & 8.3 & 1.72  & 11.8\\
Cop01r   & 1402.2  & 6.5 & 2.75  & 0.3\\
\hline
Cop11r   &  323.7  & 1.6 & 0.67  & 2.4\\
Cop11ri  &  369.7  & 1.8 & 0.38  & 1.0\\
Copal11r &  590.6  & 1.5 & 0.37  & 1.0\\
Copal11ri & 342.6  & 1.6 & 0.35  & 0.8\\
\enddata
\tablecomments{The function $\chi^{2}$ defined as $\chi^{2}=\frac{1}{N}\sum^{N}_{i=1} \frac{(q_{\rm ob, i}-q_{\rm th, i})^{2}}{\sigma^{2}_{i}}$,
where $q_{\rm ob,i}$ and $q_{\rm th,i}$ are the observed/inferred and theoretical values of quantities $q_{\rm i}$, respectively;
$\sigma_{i}$ are the errors associated to the corresponding observed/inferred quantities; $N$ is the number of the quantities.
\tablenotetext{a}{The inferred $c_{s}$ and $\rho$ are given in \citet{basu09}.}
\tablenotetext{b}{The observed $d_{02}$ and $d_{13}$ are calculated from the frequencies given by \citet{cha99b}.}
\tablenotetext{c}{In order to calculate the value of $\chi^{2}_{\rm Borexino-neutrino}$, the $\Phi(hep)<220\times10^{3}$
$\mathrm{cm}^{2}$ $\mathrm{s}^{-1}$ of \citet{bore18} was replaced by $\Phi(hep)=19^{+12}_{-9}\times10^{3}$ $\mathrm{cm}^{2}$ $\mathrm{s}^{-1}$
of \citet{berg16}.}
\tablenotetext{d}{The inferred helium is $0.2485\pm0.0035$ \citep{basu04}.}
}
\end{deluxetable*}

\clearpage

\appendix
\section{Opacity Tables}
\label{appda}
The low-temperature opacity tables with the mixtures listed in Table \ref{tabop} were downloaded from
\url{https://www.wichita.edu/academics/fairmount\_college\_of\_liberal\_arts\_and\_sciences/physics/Research/opacity.php}.
The OPAL and OP opacity tables with the same mixtures were computed from \url{https://opalopacity.llnl.gov/type1inp.html}
and \url{http://op-opacity.obspm.fr/opacity/}, respectively. The OP opacities did not consider the
contributions of P, Cl, K, and Ti compared to OPAL opacities. The contributions of the missing
elements (P, Cl, K, and Ti) were automatically transferred to other elements (see the note of
the website of OP). The opacity tables, used in this work, are also available at the author's
github repository at \url{https://github.com/yangwuming/SUN}.

The low-temperature opacity tables were used in the low-temperature region with $\lg T<4.0$,
and OPAL or OP tables were used in the high-temperature region with $\lg T>4.1$.
In the region with $4.0 \leq \lg T \leq 4.1$, the opacity was linearly interpolated between
the high-temperature and low-temperature opacities. Moreover, we corrected the wrong CALL
YALO2D in the subroutine YALO3D2 of YREC, which mainly affects the value of mixing-length parameter
\dalpha{} (lead to a small decrease in \dalpha{}). For models Aop12ri, Copal11ri, and Cop11ri, we applied
a straight multiplier $k_{0}$ to opacity $\kappa$ to enhance the opacity in a given region (see Section \ref{inopacity}).

\begin{deluxetable*}{ccc}
% \tablenum{A1}
\tablecaption{ Fractional Abundances of Heavy Elements Used to Construct Opacity Tables.
\label{tabop}}
% \tablewidth{0pt}
\tablehead{
 Element & Number fraction & Number fraction \\
         & (Caffau's mixtures) & (AAG21 mixtures)
}
\startdata
C	& 	0.260408 & 0.266509  \\
N	& 	0.059656 & 0.062476  \\
O	& 	0.473865 & 0.452597  \\
Ne	&	0.096751 & 0.106099  \\
Na	&	0.001681 & 0.001534  \\
Mg	&	0.029899 & 0.032788  \\
Al	&	0.002487 & 0.002487  \\
Si	&	0.029218 & 0.029903  \\
P	&	0.000237 & 0.000238  \\
S	&	0.011632 & 0.012182  \\
Cl	&	0.000150 & 0.000189  \\
Ar	&	0.002727 & 0.002217  \\
K	&	0.000106 & 0.000109  \\
Ca	&	0.001760 & 0.001844  \\
Ti	&	0.000072 & 0.000086  \\
Cr	&	0.000385 & 0.000385  \\
Mn	&	0.000266 & 0.000243  \\
Fe	&	0.027268 & 0.026651  \\
Ni	&	0.001431 & 0.001465  \\
\enddata
\tablecomments{The number fraction of Caffau's mixtures were listed by Ferguson in the WSU low
temperature opacities, while those of AAG21 mixtures were calculated by using the data in Table
2 of \citet{aspl21}. }
\end{deluxetable*}

\section{Nuclear reactions}
\label{appdb}
The nuclear reaction rates in conjunction with the corresponding energy release of three alternative
$pp$ branches ($pp1, pp2, pp3$), CNO cycle, and helium burning are calculated in YREC. The sequence of the
reactions is explicitly listed in \citet{dema08}. The standard reaction rates implemented are identical to
the rates published in \citet{bah89}. The calculated energy generation and neutrino fluxes are dependent on
Q-values and nuclear cross-section factors $S_{0}$, respectively. The default Q-values are taken from the Table 21
of \citet{bah88} and not changed. The default values of nuclear cross-section factors $S_{0}$ are taken from
the Tables 3.2 and 3.4 of \citet{bah89}. Compared with the default and those adopted by \citet{bah92}, we adopted
different $S_{0}(hep)$ and $S_{0}(^{7}\rm{Be}+p)$, which are shown in Table \ref{tabnuc}. The value of $0.0202$
keV barns of $S_{0}(^{7}\rm{Be}+p)$ \citep{schr94} is in agreement with $0.0205\pm0.0007$ keV barns of \citet{baby03}
and $0.0208\pm0.0007$ keV barns of \citet{adel11}. Moreover, if $S_{0}(hep)=15.3\times10^{-20}$ keV barns \citep{wolf89}
is adopted, the value of $hep$ flux of model Copal11r is $14.4\times10^{3}$ cm$^{2}$ s$^{-1}$, which is in better
agreement with that determined by \citet{berg16}. The changes in $S_{0}(hep)$ and $S_{0}(^{7}\rm{Be}+p)$ affect,
respectively, only the $hep$ and $^{8}$B fluxes \citep{bah04a}. All models were computed by using the same Q-values
and $S_{0}$.

\begin{deluxetable*}{lllll}
% \tablenum{A2}
\tablecaption{Some of Nuclear Cross-section Factors $S_{0}$ (keV barns).
\label{tabnuc}}
\tablewidth{0pt}
\tablehead{
 Reaction & Default$^{a}$ & BP92 & BP04 & This work
}
\startdata
  $^{1}$H($p$, e$^{+}\nu_{e}$ ) $^{2}$H  & $4.07\times10^{-22}$ & $4.00^{+0.06}_{-0.04}\times10^{-22}$ & $3.94(1\pm0.004)\times10^{-22}$  & $4.00\times10^{-22}$ \\
  $^{1}$H($p+e^{-}$, $\nu_{e}$) $^{2}$H  & Eq. (3.17)$^{a}$     & Eq. (3.17)$^{a}$     & Eq. (3.17)$^{a}$      &   Eq. (3.17)$^{a}$ \\
  $^{3}$He($^{3}$He, $2p$) $^{4}$He      & $5.15\times10^{3}$   & $5.00(1\pm0.06)\times10^{3}$   & $5.4\times10^{3}$  &  $5.00\times10^{3}$\\
  $^{3}$He($^{4}$He, $\gamma$) $^{7}$Be  & 0.54                 & $0.533(1\pm0.032)$         & 0.533  & 0.533 \\
  $^{7}$Be($e^{-}$, $\nu_{e}$) $^{7}$Li  & Eq. (3.18)$^{a}$     &  Eq. (3.18)$^{a}$    & Eq. (26)$^{b}$ & Eq. (3.18)$^{a}$ \\
  $^{7}$Be($p$, $\gamma$) $^{8}$B        & 0.0243               & $0.0224(1\pm0.093)$      & 0.0206$\pm0.0008^{c}$ &  0.0202$^{d}$ \\
  $^{3}$He($p$, e$^{+}\nu_{e}$ )$^{4}$He & $8\times10^{-20}$    & $1.30\times10^{-20}$ & $(8.6\pm1.3)\times10^{-20}$ & $2.30\times4.5^{e}\times10^{-20}$ \\
\enddata
\tablecomments{$^{a}$\citet{bah89}. $^{b}$The equation of \citet{adel98}. $^{c}S_{20\ \rm{keV}}(^{7}\rm{Be}+p)$ given by \citet{jun03}.
$^{d}$The value given by \citet{schr94}. $^{e}$The value of $S_{0}(hep)$ given by \citet{schi92} is $2.30\times10^{-20}$,
which should be multiplied by a factor of $4.5$ \citep{marc20}.
}
\end{deluxetable*}

\section{Numerical Solar Model and Data Availability}
\label{appdc}
Table \ref{tabmod} presents a basic numerical description of model Copal11r. All models and data used in this work
are availabe at the author's github repository at \url{https://github.com/yangwuming/SUN}. In the old YREC,
opacity was calculated by linearly interpolating between the opacity of a given $Z_{1}$ and that of another given $Z_{2}$ for
heavy-element diffusion. We modified the code to linearly interpolate between a series of $Z$ values (with $\delta Z=0.001$)
for the heavy-element diffusion. The amended YREC package, including pulsation code, are available from the author
at yangwuming@bnu.edu.cn.

\begin{longrotatetable}
\begin{deluxetable*}{lllllllllll}
% \tablenum{A3}
\tablecaption{Numerical Description of Model Copal11r.
\label{tabmod}}
\tablewidth{0pt}
\tablehead{
\colhead{M/M$_{\odot}$} & \colhead{r/R$_{\odot}$} & \colhead{L/L$_{\odot}$} & \colhead{T (K)} & \colhead{$\rho$ (g cm$^{-3}$)} & \colhead{p (dyn cm$^{-2}$)} & \colhead{$\kappa$ (cm$^{2}$ g$^{-1}$)} & \colhead{X} & \colhead{Y} & \colhead{Z} & \colhead{$\Omega$ (10$^{-6}$ rad s$^{-1}$)}
}
\startdata
 0.000001 & 0.00190 & 0.00001 & 1.576e+07 & 1.542e+02 & 2.368e+17 & 1.2914e+00 & 0.3364 & 0.6444 & 0.0192 & 1.50e-05 \\
 0.000002 & 0.00263 & 0.00002 & 1.576e+07 & 1.542e+02 & 2.367e+17 & 1.2916e+00 & 0.3365 & 0.6443 & 0.0192 & 1.50e-05 \\
 0.000003 & 0.00310 & 0.00003 & 1.576e+07 & 1.541e+02 & 2.367e+17 & 1.2917e+00 & 0.3366 & 0.6442 & 0.0192 & 1.50e-05 \\
 0.000005 & 0.00365 & 0.00005 & 1.576e+07 & 1.541e+02 & 2.366e+17 & 1.2918e+00 & 0.3367 & 0.6441 & 0.0192 & 1.50e-05 \\
 0.000009 & 0.00430 & 0.00008 & 1.576e+07 & 1.540e+02 & 2.366e+17 & 1.2920e+00 & 0.3369 & 0.6439 & 0.0192 & 1.50e-05 \\
 0.000014 & 0.00507 & 0.00013 & 1.575e+07 & 1.539e+02 & 2.364e+17 & 1.2922e+00 & 0.3371 & 0.6437 & 0.0192 & 1.50e-05 \\
 0.000023 & 0.00597 & 0.00021 & 1.575e+07 & 1.538e+02 & 2.363e+17 & 1.2926e+00 & 0.3375 & 0.6434 & 0.0192 & 1.50e-05 \\
 0.000038 & 0.00704 & 0.00034 & 1.574e+07 & 1.537e+02 & 2.361e+17 & 1.2931e+00 & 0.3379 & 0.6430 & 0.0191 & 1.50e-05 \\
 0.000062 & 0.00829 & 0.00055 & 1.574e+07 & 1.534e+02 & 2.357e+17 & 1.2938e+00 & 0.3385 & 0.6424 & 0.0191 & 1.50e-05 \\
 0.000102 & 0.00977 & 0.00089 & 1.573e+07 & 1.531e+02 & 2.353e+17 & 1.2947e+00 & 0.3393 & 0.6415 & 0.0191 & 1.50e-05 \\
 0.000166 & 0.01152 & 0.00146 & 1.571e+07 & 1.527e+02 & 2.347e+17 & 1.2960e+00 & 0.3405 & 0.6403 & 0.0191 & 1.50e-05 \\
 0.000272 & 0.01358 & 0.00238 & 1.570e+07 & 1.521e+02 & 2.339e+17 & 1.2979e+00 & 0.3421 & 0.6387 & 0.0191 & 1.50e-05 \\
 0.000444 & 0.01601 & 0.00387 & 1.567e+07 & 1.512e+02 & 2.328e+17 & 1.3004e+00 & 0.3444 & 0.6365 & 0.0191 & 1.50e-05 \\
 0.000726 & 0.01889 & 0.00629 & 1.564e+07 & 1.501e+02 & 2.312e+17 & 1.3039e+00 & 0.3475 & 0.6334 & 0.0191 & 1.49e-05 \\
 0.001187 & 0.02230 & 0.01022 & 1.559e+07 & 1.485e+02 & 2.291e+17 & 1.3087e+00 & 0.3517 & 0.6292 & 0.0191 & 1.49e-05 \\
 0.001939 & 0.02634 & 0.01655 & 1.552e+07 & 1.463e+02 & 2.261e+17 & 1.3149e+00 & 0.3576 & 0.6233 & 0.0191 & 1.48e-05 \\
 0.003168 & 0.03116 & 0.02669 & 1.543e+07 & 1.434e+02 & 2.221e+17 & 1.3236e+00 & 0.3656 & 0.6153 & 0.0191 & 1.48e-05 \\
 0.005179 & 0.03691 & 0.04284 & 1.530e+07 & 1.394e+02 & 2.166e+17 & 1.3357e+00 & 0.3766 & 0.6044 & 0.0191 & 1.47e-05 \\
 0.008465 & 0.04382 & 0.06827 & 1.512e+07 & 1.342e+02 & 2.092e+17 & 1.3526e+00 & 0.3914 & 0.5895 & 0.0190 & 1.45e-05 \\
 0.013830 & 0.05217 & 0.10770 & 1.488e+07 & 1.274e+02 & 1.993e+17 & 1.3763e+00 & 0.4114 & 0.5696 & 0.0190 & 1.43e-05 \\
 0.022608 & 0.06237 & 0.16760 & 1.455e+07 & 1.187e+02 & 1.863e+17 & 1.4098e+00 & 0.4378 & 0.5433 & 0.0189 & 1.41e-05 \\
 0.036949 & 0.07497 & 0.25560 & 1.410e+07 & 1.079e+02 & 1.694e+17 & 1.4577e+00 & 0.4720 & 0.5092 & 0.0188 & 1.37e-05 \\
 0.056968 & 0.08871 & 0.36217 & 1.358e+07 & 9.657e+01 & 1.509e+17 & 1.5169e+00 & 0.5091 & 0.4722 & 0.0187 & 1.33e-05 \\
 0.080715 & 0.10213 & 0.46885 & 1.304e+07 & 8.624e+01 & 1.333e+17 & 1.5799e+00 & 0.5433 & 0.4380 & 0.0186 & 1.30e-05 \\
 0.109607 & 0.11617 & 0.57552 & 1.248e+07 & 7.636e+01 & 1.161e+17 & 1.6526e+00 & 0.5758 & 0.4057 & 0.0185 & 1.26e-05 \\
 0.146362 & 0.13193 & 0.68219 & 1.185e+07 & 6.642e+01 & 9.836e+16 & 1.7429e+00 & 0.6072 & 0.3744 & 0.0184 & 1.22e-05 \\
 0.196727 & 0.15132 & 0.78841 & 1.110e+07 & 5.573e+01 & 7.918e+16 & 1.8647e+00 & 0.6383 & 0.3434 & 0.0183 & 1.18e-05 \\
 0.254970 & 0.17201 & 0.87065 & 1.033e+07 & 4.595e+01 & 6.194e+16 & 2.0069e+00 & 0.6631 & 0.3188 & 0.0182 & 1.14e-05 \\
 0.312918 & 0.19172 & 0.92300 & 9.650e+06 & 3.798e+01 & 4.845e+16 & 2.1549e+00 & 0.6810 & 0.3010 & 0.0181 & 1.04e-05 \\
 0.369383 & 0.21066 & 0.95531 & 9.044e+06 & 3.149e+01 & 3.790e+16 & 2.3107e+00 & 0.6909 & 0.2911 & 0.0180 & 9.57e-06 \\
 0.423543 & 0.22902 & 0.97493 & 8.501e+06 & 2.610e+01 & 2.965e+16 & 2.4761e+00 & 0.6975 & 0.2846 & 0.0179 & 9.20e-06 \\
 0.474881 & 0.24694 & 0.98679 & 8.011e+06 & 2.160e+01 & 2.319e+16 & 2.6534e+00 & 0.7022 & 0.2800 & 0.0179 & 8.80e-06 \\
 0.523093 & 0.26456 & 0.99366 & 7.567e+06 & 1.786e+01 & 1.814e+16 & 2.8464e+00 & 0.7054 & 0.2768 & 0.0178 & 8.40e-06 \\
 0.568053 & 0.28198 & 0.99705 & 7.162e+06 & 1.474e+01 & 1.419e+16 & 3.0574e+00 & 0.7078 & 0.2745 & 0.0177 & 8.03e-06 \\
 0.609743 & 0.29929 & 0.99852 & 6.789e+06 & 1.215e+01 & 1.110e+16 & 3.2874e+00 & 0.7098 & 0.2725 & 0.0177 & 7.66e-06 \\
 0.648210 & 0.31657 & 0.99923 & 6.446e+06 & 1.001e+01 & 8.682e+15 & 3.5379e+00 & 0.7115 & 0.2709 & 0.0177 & 7.31e-06 \\
 0.683576 & 0.33387 & 0.99960 & 6.126e+06 & 8.229e+00 & 6.792e+15 & 3.8109e+00 & 0.7129 & 0.2695 & 0.0176 & 6.96e-06 \\
 0.715985 & 0.35125 & 0.99980 & 5.829e+06 & 6.762e+00 & 5.312e+15 & 4.1078e+00 & 0.7141 & 0.2683 & 0.0176 & 6.64e-06 \\
 0.745591 & 0.36875 & 0.99991 & 5.550e+06 & 5.552e+00 & 4.156e+15 & 4.4296e+00 & 0.7151 & 0.2673 & 0.0176 & 6.32e-06 \\
 0.772576 & 0.38642 & 0.99997 & 5.287e+06 & 4.556e+00 & 3.251e+15 & 4.7782e+00 & 0.7161 & 0.2664 & 0.0175 & 6.02e-06 \\
 0.797116 & 0.40428 & 1.00000 & 5.040e+06 & 3.737e+00 & 2.543e+15 & 5.1556e+00 & 0.7169 & 0.2656 & 0.0175 & 5.74e-06 \\
 0.819381 & 0.42237 & 1.00001 & 4.806e+06 & 3.064e+00 & 1.989e+15 & 5.5644e+00 & 0.7177 & 0.2648 & 0.0175 & 5.46e-06 \\
 0.839548 & 0.44069 & 1.00001 & 4.584e+06 & 2.511e+00 & 1.556e+15 & 6.0050e+00 & 0.7186 & 0.2640 & 0.0174 & 5.21e-06 \\
 0.857782 & 0.45929 & 1.00001 & 4.373e+06 & 2.058e+00 & 1.217e+15 & 6.4815e+00 & 0.7194 & 0.2632 & 0.0174 & 4.96e-06 \\
 0.874233 & 0.47816 & 1.00000 & 4.172e+06 & 1.686e+00 & 9.520e+14 & 7.0002e+00 & 0.7203 & 0.2624 & 0.0174 & 4.73e-06 \\
 0.889056 & 0.49732 & 1.00000 & 3.980e+06 & 1.382e+00 & 7.447e+14 & 7.5656e+00 & 0.7212 & 0.2614 & 0.0174 & 4.51e-06 \\
 0.902387 & 0.51678 & 0.99999 & 3.795e+06 & 1.132e+00 & 5.825e+14 & 8.1805e+00 & 0.7223 & 0.2604 & 0.0173 & 4.30e-06 \\
 0.914354 & 0.53654 & 0.99998 & 3.618e+06 & 9.283e-01 & 4.557e+14 & 8.8490e+00 & 0.7234 & 0.2593 & 0.0173 & 4.11e-06 \\
 0.925080 & 0.55659 & 0.99998 & 3.446e+06 & 7.614e-01 & 3.564e+14 & 9.5772e+00 & 0.7247 & 0.2580 & 0.0173 & 3.92e-06 \\
 0.934675 & 0.57691 & 0.99997 & 3.280e+06 & 6.251e-01 & 2.788e+14 & 1.0375e+01 & 0.7262 & 0.2565 & 0.0173 & 3.75e-06 \\
 0.943239 & 0.59749 & 0.99997 & 3.118e+06 & 5.137e-01 & 2.181e+14 & 1.1257e+01 & 0.7278 & 0.2549 & 0.0173 & 3.58e-06 \\
 0.950869 & 0.61828 & 0.99996 & 2.958e+06 & 4.230e-01 & 1.706e+14 & 1.2250e+01 & 0.7296 & 0.2531 & 0.0173 & 3.42e-06 \\
 0.957649 & 0.63923 & 0.99996 & 2.799e+06 & 3.491e-01 & 1.334e+14 & 1.3427e+01 & 0.7316 & 0.2510 & 0.0174 & 3.28e-06 \\
 0.963657 & 0.66025 & 0.99997 & 2.637e+06 & 2.894e-01 & 1.044e+14 & 1.5065e+01 & 0.7338 & 0.2485 & 0.0177 & 3.14e-06 \\
 0.968963 & 0.68119 & 0.99997 & 2.467e+06 & 2.417e-01 & 8.165e+13 & 1.6387e+01 & 0.7360 & 0.2467 & 0.0173 & 3.02e-06 \\
 0.973526 & 0.70160 & 0.99998 & 2.295e+06 & 2.041e-01 & 6.424e+13 & 1.7759e+01 & 0.7382 & 0.2454 & 0.0164 & 2.92e-06 \\
 0.976002 & 0.71333 & 1.00022 & 2.185e+06 & 1.854e-01 & 5.560e+13 & 1.8426e+01 & 0.7396 & 0.2450 & 0.0155 & 2.87e-06 \\
 0.977337 & 0.72008 & 1.00037 & 2.114e+06 & 1.764e-01 & 5.118e+13 & 1.9703e+01 & 0.7396 & 0.2450 & 0.0155 & 2.87e-06 \\
 0.980936 & 0.73943 & 1.00030 & 1.918e+06 & 1.522e-01 & 4.004e+13 & 2.3806e+01 & 0.7396 & 0.2450 & 0.0155 & 2.87e-06 \\
 0.984045 & 0.75784 & 1.00024 & 1.741e+06 & 1.313e-01 & 3.132e+13 & 2.8373e+01 & 0.7396 & 0.2450 & 0.0155 & 2.87e-06 \\
 0.986711 & 0.77528 & 1.00019 & 1.579e+06 & 1.133e-01 & 2.450e+13 & 3.3468e+01 & 0.7396 & 0.2450 & 0.0155 & 2.87e-06 \\
 0.988982 & 0.79177 & 1.00016 & 1.433e+06 & 9.775e-02 & 1.916e+13 & 3.9246e+01 & 0.7396 & 0.2450 & 0.0155 & 2.87e-06 \\
 0.990904 & 0.80729 & 1.00013 & 1.300e+06 & 8.435e-02 & 1.499e+13 & 4.5830e+01 & 0.7396 & 0.2450 & 0.0155 & 2.87e-06 \\
 0.992521 & 0.82188 & 1.00011 & 1.180e+06 & 7.278e-02 & 1.173e+13 & 5.3219e+01 & 0.7396 & 0.2450 & 0.0155 & 2.87e-06 \\
 0.993874 & 0.83555 & 1.00009 & 1.070e+06 & 6.280e-02 & 9.173e+12 & 6.1516e+01 & 0.7396 & 0.2450 & 0.0155 & 2.87e-06 \\
 0.995001 & 0.84832 & 1.00008 & 9.710e+05 & 5.419e-02 & 7.175e+12 & 7.0972e+01 & 0.7396 & 0.2450 & 0.0155 & 2.87e-06 \\
 0.995933 & 0.86022 & 1.00007 & 8.811e+05 & 4.677e-02 & 5.613e+12 & 8.2430e+01 & 0.7396 & 0.2450 & 0.0155 & 2.87e-06 \\
 0.996702 & 0.87130 & 1.00006 & 7.995e+05 & 4.036e-02 & 4.390e+12 & 9.6834e+01 & 0.7396 & 0.2450 & 0.0155 & 2.87e-06 \\
 0.997334 & 0.88158 & 1.00006 & 7.254e+05 & 3.483e-02 & 3.434e+12 & 1.1441e+02 & 0.7396 & 0.2450 & 0.0155 & 2.87e-06 \\
 0.997850 & 0.89110 & 1.00005 & 6.583e+05 & 3.006e-02 & 2.686e+12 & 1.3527e+02 & 0.7396 & 0.2450 & 0.0155 & 2.87e-06 \\
 0.998271 & 0.89991 & 1.00005 & 5.975e+05 & 2.594e-02 & 2.101e+12 & 1.5967e+02 & 0.7396 & 0.2450 & 0.0155 & 2.87e-06 \\
 0.998613 & 0.90804 & 1.00005 & 5.423e+05 & 2.239e-02 & 1.644e+12 & 1.8885e+02 & 0.7396 & 0.2450 & 0.0155 & 2.87e-06 \\
 0.998889 & 0.91554 & 1.00005 & 4.923e+05 & 1.932e-02 & 1.286e+12 & 2.2482e+02 & 0.7396 & 0.2450 & 0.0155 & 2.87e-06 \\
 0.999113 & 0.92244 & 1.00005 & 4.469e+05 & 1.668e-02 & 1.006e+12 & 2.7065e+02 & 0.7396 & 0.2450 & 0.0155 & 2.87e-06 \\
 0.999292 & 0.92878 & 1.00005 & 4.058e+05 & 1.439e-02 & 7.867e+11 & 3.3250e+02 & 0.7396 & 0.2450 & 0.0155 & 2.87e-06 \\
 0.999436 & 0.93461 & 1.00005 & 3.685e+05 & 1.242e-02 & 6.154e+11 & 4.1670e+02 & 0.7396 & 0.2450 & 0.0155 & 2.87e-06 \\
 0.999552 & 0.93995 & 1.00004 & 3.348e+05 & 1.072e-02 & 4.814e+11 & 5.3263e+02 & 0.7396 & 0.2450 & 0.0155 & 2.87e-06 \\
 0.999644 & 0.94484 & 1.00004 & 3.042e+05 & 9.247e-03 & 3.765e+11 & 6.9405e+02 & 0.7396 & 0.2450 & 0.0155 & 2.87e-06 \\
 0.999718 & 0.94933 & 1.00004 & 2.765e+05 & 7.979e-03 & 2.946e+11 & 9.2024e+02 & 0.7396 & 0.2450 & 0.0155 & 2.87e-06 \\
 0.999777 & 0.95343 & 1.00004 & 2.514e+05 & 6.883e-03 & 2.304e+11 & 1.2436e+03 & 0.7396 & 0.2450 & 0.0155 & 2.87e-06 \\
 0.999823 & 0.95718 & 1.00004 & 2.288e+05 & 5.936e-03 & 1.802e+11 & 1.7082e+03 & 0.7396 & 0.2450 & 0.0155 & 2.87e-06 \\
 0.999860 & 0.96060 & 1.00004 & 2.083e+05 & 5.118e-03 & 1.410e+11 & 2.3811e+03 & 0.7396 & 0.2450 & 0.0155 & 2.87e-06 \\
 0.999890 & 0.96373 & 1.00004 & 1.898e+05 & 4.411e-03 & 1.103e+11 & 3.3552e+03 & 0.7396 & 0.2450 & 0.0155 & 2.87e-06 \\
 0.999913 & 0.96659 & 1.00004 & 1.731e+05 & 3.798e-03 & 8.626e+10 & 4.7501e+03 & 0.7396 & 0.2450 & 0.0155 & 2.87e-06 \\
 0.999931 & 0.96921 & 1.00004 & 1.581e+05 & 3.269e-03 & 6.748e+10 & 6.6841e+03 & 0.7396 & 0.2450 & 0.0155 & 2.87e-06 \\
 0.999946 & 0.97159 & 1.00004 & 1.447e+05 & 2.810e-03 & 5.278e+10 & 9.2032e+03 & 0.7396 & 0.2450 & 0.0155 & 2.87e-06 \\
 0.999957 & 0.97378 & 1.00004 & 1.326e+05 & 2.412e-03 & 4.129e+10 & 1.2213e+04 & 0.7396 & 0.2450 & 0.0155 & 2.87e-06 \\
 0.999967 & 0.97578 & 1.00004 & 1.217e+05 & 2.068e-03 & 3.230e+10 & 1.5468e+04 & 0.7396 & 0.2450 & 0.0155 & 2.87e-06 \\
 0.999974 & 0.97761 & 1.00004 & 1.118e+05 & 1.772e-03 & 2.526e+10 & 1.8710e+04 & 0.7396 & 0.2450 & 0.0155 & 2.87e-06 \\
 0.999979 & 0.97930 & 1.00004 & 1.028e+05 & 1.516e-03 & 1.976e+10 & 2.1823e+04 & 0.7396 & 0.2450 & 0.0155 & 2.87e-06 \\
 0.999984 & 0.98084 & 1.00004 & 9.461e+04 & 1.298e-03 & 1.546e+10 & 2.4907e+04 & 0.7396 & 0.2450 & 0.0155 & 2.87e-06 \\
 0.999987 & 0.98225 & 1.00004 & 8.699e+04 & 1.111e-03 & 1.209e+10 & 2.8321e+04 & 0.7396 & 0.2450 & 0.0155 & 2.87e-06 \\
 0.999990 & 0.98355 & 1.00004 & 7.994e+04 & 9.521e-04 & 9.458e+09 & 3.2562e+04 & 0.7396 & 0.2450 & 0.0155 & 2.87e-06 \\
 0.999994 & 0.98581 & 1.00004 & 6.744e+04 & 7.004e-04 & 5.788e+09 & 4.5889e+04 & 0.7396 & 0.2450 & 0.0155 & 2.87e-06 \\
 0.999996 & 0.98771 & 1.00004 & 5.714e+04 & 5.143e-04 & 3.541e+09 & 6.9908e+04 & 0.7396 & 0.2450 & 0.0155 & 2.87e-06 \\
 0.999998 & 0.98929 & 1.00004 & 4.894e+04 & 3.749e-04 & 2.167e+09 & 1.0278e+05 & 0.7396 & 0.2450 & 0.0155 & 2.87e-06 \\
 0.999999 & 0.99064 & 1.00004 & 4.252e+04 & 2.701e-04 & 1.326e+09 & 1.2664e+05 & 0.7396 & 0.2450 & 0.0155 & 2.87e-06 \\
 0.999999 & 0.99179 & 1.00004 & 3.746e+04 & 1.923e-04 & 8.113e+08 & 1.2530e+05 & 0.7396 & 0.2450 & 0.0155 & 2.87e-06 \\
 0.999999 & 0.99231 & 1.00004 & 3.531e+04 & 1.616e-04 & 6.346e+08 & 1.1637e+05 & 0.7396 & 0.2450 & 0.0155 & 2.87e-06 \\
 0.999999 & 0.99279 & 1.00004 & 3.337e+04 & 1.355e-04 & 4.964e+08 & 1.0456e+05 & 0.7396 & 0.2450 & 0.0155 & 2.87e-06 \\
 1.000000 & 0.99354 & 1.00004 & 3.044e+04 & 9.966e-05 & 3.255e+08 & 8.2105e+04 & 0.7396 & 0.2450 & 0.0155 & 2.87e-06 \\
 1.000000 & 0.99444 & 1.00004 & 2.716e+04 & 6.534e-05 & 1.845e+08 & 5.6764e+04 & 0.7396 & 0.2450 & 0.0155 & 2.87e-06 \\
 1.000000 & 0.99594 & 1.00004 & 2.214e+04 & 2.646e-05 & 5.699e+07 & 2.3956e+04 & 0.7396 & 0.2450 & 0.0155 & 2.87e-06 \\
 1.000000 & 0.99786 & 1.00004 & 1.643e+04 & 4.852e-06 & 6.879e+06 & 4.6756e+03 & 0.7396 & 0.2450 & 0.0155 & 2.87e-06 \\
 1.000000 & 0.99837 & 1.00004 & 1.495e+04 & 2.667e-06 & 3.292e+06 & 2.5776e+03 & 0.7396 & 0.2450 & 0.0155 & 2.87e-06 \\
 1.000000 & 0.99881 & 1.00004 & 1.360e+04 & 1.469e-06 & 1.576e+06 & 1.3462e+03 & 0.7396 & 0.2450 & 0.0155 & 2.87e-06 \\
 1.000000 & 0.99919 & 1.00004 & 1.230e+04 & 8.173e-07 & 7.542e+05 & 6.2276e+02 & 0.7396 & 0.2450 & 0.0155 & 2.87e-06 \\
 1.000000 & 0.99952 & 1.00004 & 1.090e+04 & 4.663e-07 & 3.610e+05 & 2.0916e+02 & 0.7396 & 0.2450 & 0.0155 & 2.87e-06 \\
 1.000000 & 0.99979 & 1.00004 & 8.935e+03 & 2.874e-07 & 1.728e+05 & 2.7659e+01 & 0.7396 & 0.2450 & 0.0155 & 2.87e-06 \\
 1.000000 & 0.99998 & 1.00004 & 6.028e+03 & 2.073e-07 & 8.270e+04 & 5.0052e-01 & 0.7396 & 0.2450 & 0.0155 & 2.87e-06 \\
 1.000000 & 1.00002 & 1.00004 & 5.777e+03 & 1.719e-07 & 6.569e+04 & 3.1819e-01 & 0.7396 & 0.2450 & 0.0155 & 2.87e-06 \\
\enddata
% \tablecomments{xxx.}
\end{deluxetable*}
\end{longrotatetable}


\begin{thebibliography}{}
\bibitem[Adelberger et al.(1998)]{adel98} Adelberger, E. G., et al. 1998, Rev. Mod. Phys., 70, 1265
\bibitem[Adelberger et al.(2011)]{adel11} Adelberger E. G., et al. 2011, RMP, 83, 195
\bibitem[Ahmed et al.(2004)]{ahm04} Ahmed, S. N., Anthony, A. E., Beier, E. W. et al. 2004, PhRvL, 92, 1301
\bibitem[Amarsi et al.(2021)]{amar21} Amarsi, A. M., Grevesse, N., Asplund, M., Collet, R. 2021, A\&A, 656, 113
\bibitem[Antia \& Basu(2006)]{anti06} Antia, H. M., \& Basu, S. 2006, ApJ, 644, 1292
\bibitem[Asplund et al.(2021)]{aspl21} Asplund, M., Amarsi, A. M., Grevesse, N. 2021, A\&A 653, A141
\bibitem[Asplund et al.(2004)]{aspl04} Asplund, M., Grevesesse N., Sauval, A. J., Allende Prieto, C., \& Kiselman, D. 2004, A\&A, 417, 751
\bibitem[Asplund et al.(2005)]{aspl05} Asplund, M., Grevesesse, N., Sauval, A. J., Allende Prieto, C., \& Blomme, R.
2005, A\&A, 431, 693
\bibitem[Asplund et al.(2009)]{aspl09} Asplund, M., Grevesse, N., Sauval, A., \& Scott, P. 2009, ARA\&A, 47, 481 (AGSS09)

\bibitem[Baby et al.(2003)]{baby03} Baby, L. T., et al. 2003a, Phys. Rev. Lett., 90, 022501
\bibitem[Badnell et al.(2005)]{badn05} Badnell, N. R., Bautista, M. A., Butler, K., Delahaye, F.,
Mendoza, C., Palmeri, P., Zeippen, C. J., Seaton, M. J. 2005, MNRAS, 360, 458
\bibitem[Bahcall (1989)]{bah89} Bahcall, J. N. , 1989, Neutrino Astrophysics (Cambridge University, Cambridge, England).
\bibitem[Bahcall et al.(2001)]{bah01} Bahcall, J. N., Pinsonneault, M. H.,
\& Basu, S, 2001, ApJ, 555, 990
\bibitem[Bahcall \& Pinsonneault(1992)]{bah92} Bahcall, J. N., \& Pinsonneault, M. H.
1992, Rev. Mod. Phys., 64, 885 (BP92)
\bibitem[Bahcall et al.(1995)]{bah95} Bahcall J. N., Pinsonneault M. H.,
\& Wasserburg G. J. 1995, Rev. Mod. Phys., 67, 781
\bibitem[Bahcall et al.(2005)]{bah05} Bahcall, J. N., Basu, S., Pinsonneault, M. H., \& Serenelli, A. M. 2005, ApJ,
618, 1049
\bibitem[Bahcall \& Pinsonneault(2004)]{bah04a} Bahcall, J. N., \& Pinsonneault, M. H. 2004, Phys. Rev. Lett., 92, 121301 (BP04)
\bibitem[Bahcall et al.(2004)]{bah04b} Bahcall, J. N., Serenelli, A. M., \& Pinsonneault, M. H. 2004, ApJ, 614, 464
%\bibitem[Bahcall et al.(2006)]{bah06} Bahcall, J. N., Serenelli, A. M., \& Basu, S. 2006, ApJS, 165, 400
\bibitem[Bahcall \& Ulrich(1988)]{bah88} Bahcall, J. N., \& Ulrich, R. 1988, RvMP, 60, 297
\bibitem[Ball(2022)]{ball22} Ball, W. H. 2021, RNAAS, 5, 7
\bibitem[Basu \& Antia(1997)]{basu97} Basu, S., \& Antia, H. M. 1997, MNRAS, 287, 189
\bibitem[Basu \& Antia(2004)]{basu04} Basu, S., \& Antia, H. M. 2004, ApJL, 606, L85
\bibitem[Basu et al.(2009)]{basu09} Basu, S., Chaplin, W. J., Elsworth, Y., New, R., \& Serenelli, A. M. 2009, ApJ, 699, 1403
\bibitem[Basu et al.(2015)]{basu15} Basu, S., Grevesse, N., Mathis, S., Turck-Chi\`{e}ze, S. 2015, Space Sci Rev., 196, 49
\bibitem[Basu et al.(2000)]{basu00} Basu, S., Pinsonneault, M. H., \& Bahcall, J. N. 2000, ApJ, 529, 1084
\bibitem[Bellini et al.(2011)]{bel11} Bellini, G. Benziger, J. Bick, D. et al. 2011, PhRvL, 107, 141302
\bibitem[Bellini et al.(2012)]{bel12} Bellini, G. Benziger, J. Bick, D. et al. 2012, PhRvL, 108, 51302

\bibitem[Bergstr\"{o}m et al.(2016)]{berg16} Bergstr\"{o}m, J., Gonzalez-Garcia, M. C., Maltoni, M., et al. 2016, JHEP, 3, 132
\bibitem[B\"{o}hm-Vitense(1958)]{boh58} B\"{o}hm-Vitense, E. 1958, ZAp, 46, 108
\bibitem[Borexino Collaboration(2018)]{bore18} Borexino Collaboration, Agostini, M., Altenm\"{u}ller, K., et al. 2018, Nature, 562, 505
\bibitem[Borexino Collaboration(2020)]{bore20} Borexino Collaboration, Agostini, M., Altenm\"{u}ller, K., Appel, S., et al. 2020, Nature, 587, 577
\bibitem[Buldgen et al.(2017)]{buld17} Buldgen, G.,Salmon, S. J. A. J., Noels, A., Scuflaire, R.,
Dupret, M. A., Reese, D. R. 2017, MNRAS, 472, 751
\bibitem[Buldgen et al.(2019)]{buld19} Buldgen, G., Salmon, S. J. A. J., Noels, A., et al. 2019, A\&A, 621, 33

\bibitem[Caffau et al.(2010)]{caf10} Caffau, E., Ludwig, H.-G., Bonifacio, P., Faraggiana, R., Steffen, M., Freytag, B., Kamp, I., Ayres, T. R. 2010, A\&A, 514, A92
\bibitem[Caffau et al.(2011)]{caf11} Caffau, E., Ludwig, H.-G., Steffen, M., Freytag, B.,
Bonifacio, P. 2011, Solar Phys., 268, 255
\bibitem[Castro et al.(2007)]{cast07} Castro, M., Vauclair, S., \& Richard, O. 2007, A\&A, 463, 755
\bibitem[Chaboyer et al(1995)]{cha95} Chaboyer, B., Demarque, P., Pinsonneault, M. H. 1995, ApJ, 441, 865
\bibitem[Chaplin et al.(1996)]{cha96} Chaplin, W. J., Elsworth, Y., Howe, R., et al. 1996, Sol. Phys., 168, 1
\bibitem[Chaplin et al.(1999)]{cha99b}
Chaplin, W. J., Elsworth, Y., Isaak, G. R., Miller, B. A., \& New, R. 1999b, MNRAS, 308, 424
\bibitem[Christensen-Dalsgaard et al.(1991)]{chr91}
Christensen-Dalsgaard, J., Gough, D. O., \& Thompson, M. J. 1991, ApJ, 378, 413
\bibitem[Christensen-Dalsgaard(2021)]{chris21} Christensen-Dalsgaard, J. 2021, Living Reviews in Solar Physics, 18, 2
\bibitem[Delahaye et al.(2016)]{dela16} Delahaye, F., Zw\"{o}lf, C. M., Zeippen, C. J., Mendoza, C. 2016, JQSRT, 171, 66
\bibitem[Demarque et al.(2008)]{dema08} Demarque, P., Guenther, D. B., Li, L. H., Mazumdar, A., Straka, C. W. 2008,
Ap\&SS, 316, 31
\bibitem[Endal \& Sofia(1978)]{enda78} Endal, A. S., \& Sofia, S. 1978, ApJ, 220, 279

\bibitem[Ferguson et al.(2005)]{fer05} Ferguson, J. W., Alexander, D. R., Allard, F. et al. 2005, ApJ, 623, 585
\bibitem[Garc\'{\i}a et al.(2011)]{gar11} Garc\'{\i}a, R. A., Salabert, D., \& Ballot, J. et al. 2011, JPhCS, 271, 2049
\bibitem[Grevesse \& Sauval(1998)]{grev98} Grevesse, N., \& Sauval, A. J. 1998, in Solar Composition and Its Evolution, ed.
C. Fr\"{o}hlich et al. (Dordrecht: Kluwer), 161 (GS98)
\bibitem[Guenther(1994)]{gue94} Guenther D. B., 1994, ApJ, 422, 400
\bibitem[Guzik et al.(2005)]{guzi05} Guzik, J. A., Watson, L. S. \&
Cox, A. N. 2005, ApJ, 627, 1049
\bibitem[Guzik \& Mussack(2010)]{guzi10} Guzik, J. A., \& Mussack, K. 2010, ApJ, 713, 1108

\bibitem[Hope et al.(2020)]{hope20} Hope, C., Bergemann, M., Bitsch, B., Serenelli, A. 2020, A\&A, 641, A73
\bibitem[Iglesias \& Rogers(1996)]{igl96} Iglesias, C., Rogers, F. J. 1996, ApJ, 464, 943
\bibitem[Jermyn et al.(2022)]{jerm22} Jermyn, A. S., Anders, E. H., Lecoanet, D. \& Cantiello, M. 2022, arXiv220600011
\bibitem[Joyce \& Chaboyer(2018)]{joyc18} Joyce, M., \& Chaboyer, B. 2018, ApJ, 864, 99
\bibitem[Junghans et al.(2003)]{jun03}Junghans, A. R., et al. 2003, Phys. Rev. C, 68, 065803
\bibitem[Kawaler(1988)]{kaw88} Kawaler, S. D. 1988, ApJ, 333, 236
\bibitem[Kippenhahn et al.(2012)]{kipp12} Kippenhahn, R., Weigert, A., \& Weiss, A. 2012, Stellar Structure and Evolution,
(Heidelberg: Springer)
\bibitem[Krishna Swamy(1966)]{kris66} Krishna Swamy, K. S. 1966, ApJ, 145, 174
\bibitem[Le Pennec et al.(2015)]{lep15} Le Pennec, M., Turck-Chi\`{e}ze, S.,
Salmon, S., Blancard, C., Coss\'{e}, P., Faussurier, G., Mondet, G. 2015, ApJL, 813, L42
\bibitem[Lodders(2003)]{lod03} Lodders, K. 2003, ApJ, 591, 1220
\bibitem[Lodders et al.(2009)]{lod09} Lodders, K., Palme, H., Gail, H-P. 2009, LanB, 4, 712 (LPG09)
\bibitem[Lodders(2020)]{lod20} Lodders, K. 2020, Solar Elemental Abundances, in The
Oxford Research Encyclopedia of Planetary Science, Oxford University Press, arXiv:1912.00844
\bibitem[Marcucci et al.(2000)]{marc20}  Marcucci, L. E., Schiavilla, R., Viviani, M., Kievski, A., \& Rosati, S. 2000,
PhRvL, 84, 5959
\bibitem[Montalb\'{a}n et al.(2004)]{mont04} Montalb\'{a}n, J., Miglio, A.,
Noels, A., Grevesse, N., Di Mauro, M. P. 2004, in Helio- and
Asteroseismology: Towards a Golden Future, Proc. of the SOHO 14 /
GONG 2004 Workshop, ed. D. Danesy (ESA SP-559; Noordwijk: ESA), 574
\bibitem[Montalb\'{a}n et al.(2006)]{mont06} Montalb\'{a}n, J., Miglio, A.,
Theado, S., Noels, A., Grevesse, N. 2006, Commun. Asteroseismol., 147, 80

\bibitem[Pinsonneault et al.(1989)]{pins89} Pinsonneault, M. H., Kawaler, S. D., Sofia, S., \& Demarqure, P. 1989, ApJ, 338, 424
\bibitem[Proffitt \& Michaud(1991)]{pro91} Proffitt, C. R., \& Michaud, G. 1991, ApJ, 380, 238
\bibitem[Rogers \& Nayfonov(2002)]{rog02} Rogers, F., \& Nayfonov, A. 2002, ApJ, 576, 1064
\bibitem[Roxburgh \& Vorontsov(2003)]{rox03} Roxburgh, I. W., \& Vorontsov, S. V. 2003, A\&A, 411, 215
\bibitem[Salmon et al.(2021)]{salm21} Salmon, S.J.A.J., Buldgen, G., Noels, A., Eggenberger, P., Scuflaire, R.,
Meynet, G. 2021, A\&A, 651, A106

\bibitem[Schiavilla et al.(1994)]{schi92} Schiavilla, R., Wiringa, R. B., Pandharipande, V. R., and
Carlson, J. 1992, Phys. Rev. C, 45, 2628
\bibitem[Schramm \& Shi(1994)]{schr94} Schramm, D. N., \& Shi, X. 1994, NuPhS, 35, 321
\bibitem[Seaton(1987)]{seat87} Seaton, M. J. 1987, JPhB, 20, 6363
\bibitem[Serenelli \& Basu(2010)]{sere10} Serenelli, A., \& Basu, S. 2010, ApJ, 719, 865
\bibitem[Serenelli et al.(2009)]{sere09} Serenelli, A. M., Basu, S., Ferguson, J. W., \& Asplund, M. 2009, ApJL,
705, L123
\bibitem[Serenelli et al.(2011)]{sere11} Serenelli, A., Haxton, W. C., Pe\~{n}a-Garay, C. 2011, ApJ, 743, 24
\bibitem[Schou et al.(1998)]{sch98} Schou, J., Antia, H. M., Basu, S. et al. 1998, ApJ, 505, 390
\bibitem[Spada et al.(2018)]{spad18} Spada, F., Demarque, P., Basu, S., Tanner, J. D. 2018, ApJ, 869, 135
\bibitem[Spada et al.(2019)]{spad19} Spada, F., \& Demarque, P. 2019, MNRAS, 489, 4712
\bibitem[The Opacity Project Team(1995)]{op95} The Opacity Project Team, 1995, The Opacity Project Vol. 1,
Institute of Physics Publications, Bristol, UK
\bibitem[Thoul et al.(1994)]{tho94} Thoul, A. A., Bahcall, J. N., Loeb, A. 1994, ApJ, 421, 828
\bibitem[Turck-Chi\`{e}ze \& Couvidat(2011)]{tur11a} Turck-Chi\`{e}ze, S., \& Couvidat, S. 2011, RPPh, 74, 6901
% \bibitem[Turck-Chi\`{e}ze et al.(1988)]{tur88} Turck-Chi\`{e}ze, S., Cahen, S., and Cass\`{e}, M. 1988, ApJ, 335, 415
\bibitem[Turck-Chi\`{e}ze et al.(2010)]{tur10} Turck-Chi\`{e}ze, S., Palacios, A., Marques, J. P.,
Nghiem, P. A. P. 2010, ApJ, 715, 1539
\bibitem[Turck-Chi\`{e}ze et al.(2011)]{tur11} Turck-Chi\`{e}ze, S., Piau, L., \& Couvidat, S.
2011, ApJL, 731, L29

\bibitem[Vorontsov et al.(2013)]{voro13} Vorontsov, S. V., Baturin, V. A., Ayukov, S. V., Gryaznov, V. K. 2013, MNRAS, 430, 1636
\bibitem[Vorontsov et al.(2014)]{voro14} Vorontsov, S. V., Baturin, V. A., Ayukov, S. V., Gryaznov, V. K. 2014, MNRAS, 441, 3296
\bibitem[Wolfs et al.(1989)]{wolf89} Wolfs, F. L. H.,; Freedman, S. J., Nelson, J. E., Dewey, M. S., Greene, G. L. 1989, PhRvL, 63, 2721


\bibitem[Yang(2016)]{yang16} Yang, W. 2016, ApJ, 829, 68
\bibitem[Yang(2019)]{yang19} Yang, W., 2019, ApJ, 873, 18
\bibitem[Yang \& Bi (2007)]{yang07} Yang, W. M., \& Bi, S. L. 2007, ApJL, 658, L67
\bibitem[Zhang \& Li(2012)]{zhang12} Zhang, Q., \& Li, Y. 2012, ApJ, 746, 50
\bibitem[Zhang et al.(2019)]{zhang19} Zhang, Q. S., Li, Y., Christensen-Dalsgaard, J. 2019, ApJ, 881, 103
\bibitem[Zahn(1993)]{zahn93} Zahn, J. P. 1993, in Astrophysical Fluid Dynamics, Les Houches XLVII, ed.
J.-P. Zahn \& J. Zinn-Justin (New York: Elsevier), 561
\end{thebibliography}
\end{document}